\documentstyle[graphicx, times, amsmath, amssymb, rotating, lscape, multirow]{mn2e}
\title[A Deep Study of NGC 4051 with {\sl Chandra} and {\sl Suzaku}]
{Contemporaneous Chandra HETG and Suzaku X-ray Observations of NGC 4051}

\author[A. Lobban et al.] {
A.P. Lobban$^{1}$, J.N. Reeves$^{1}$, L. Miller$^{2}$, T.J. Turner$^{3}$, V. Braito$^{4}$, S.B. Kraemer$^{5}$, D.M. Crenshaw$^{6}$\\
$^{1}$Astrophysics Group, School of Physical and Geographical Sciences, Keele 
University, Keele, Staffordshire, ST5 5BG, U.K. \\
$^{2}$Department of Physics, University of Oxford, Denys Wilkinson Building, Keble Road, Oxford, OX1 3RH, U.K. \\
$^{3}$Department of Physics, University of Maryland Baltimore County, MD 21250, U.S.A. \\
$^{4}$Department of Physics and Astronomy, University of Leicester, University Road, Leicester, LE1 7RH, U.K. \\
$^{5}$Institute for Astrophysics and Computational Sciences, Department of Physics, The Catholic University of America, Washington, DC 20064, U.S.A. \\ 
$^{6}$Department of Physics and Astronomy, Georgia State University, Astronomy Offices, One Park Place South SE, Suite 700, Atlanta, GA 30303, U.S.A.}

\date{Accepted by MNRAS on 9 February 2011}
\pagerange{\pageref{firstpage}--\pageref{lastpage}} \pubyear{2010}

\begin{document}

\def\aap{A\&A}
\def\apj{ApJ}
\def\apjl{ApJ}
\def\mnras{MNRAS}

\maketitle

\label{firstpage}

\begin{abstract}

We present the results of a deep 300\,ks {\sl Chandra} HETG observation of the highly variable narrow-line Seyfert Type 1 galaxy NGC 4051. The HETG spectrum reveals 28 significant soft X-ray ionised lines in either emission or absorption; primarily originating from H-like and He-like K-shell transitions of O, Ne, Mg and Si (including higher order lines and strong forbidden emission lines from O\,\textsc{vii} and Ne\,\textsc{ix}) plus high ionisation L-shell transitions from Fe\,\textsc{xvii} to Fe\,\textsc{xxii} and lower ionisation inner-shell lines (e.g. O\,\textsc{vi}). Modelling the data with \textsc{xstar} requires four distinct ionisation zones for the gas, all outflowing with velocities \textless 1\,000\,km\,s$^{-1}$. A selection of the strongest emission/absorption lines appear to be resolved with FWHM of $\sim$600\,km\,s$^{-1}$. We also present the results from a quasi-simultaneous 350\,ks {\sl Suzaku} observation of NGC 4051 where the XIS spectrum reveals strong evidence for blueshifted absorption lines at $\sim$6.8 and $\sim$7.1\,keV, consistent with previous findings. Modelling with \textsc{xstar} suggests that this is the signature of a highly ionised, high velocity outflow (log\,$\xi = 4.1^{+0.2}_{-0.1}$; $v_{\rm out} \sim -0.02$$c$) which potentially may have a significant effect on the host galaxy environment via feedback. Finally, we also simultaneously model the broad-band 2008 XIS+HXD {\sl Suzaku} data with archival {\sl Suzaku} data from 2005 when the source was observed to have entered an extended period of low flux in an attempt to analyse the cause of the long-term spectral variability. We find that we can account for this by allowing for large variations in the normalisation of the intrinsic power-law component which may be interpreted as being due to significant changes in the covering fraction of a Compton-thick partial-coverer obscuring the central continuum emission.

\end{abstract}


\section{Introduction}

Recent systematic X-ray studies of AGN (Active Galactic Nuclei) with {\sl ASCA}, {\sl Chandra}, {\sl XMM-Newton} and {\sl Suzaku} have established that at least half of all active galaxies host photo-ionised ``warm'' absorbers (e.g. Reynolds \& Fabian 1995; Blustin et al. 2005) and that they may in fact be ubiquitous in AGN, becoming observable only in certain lines of sight (Krongold et al. 2008; although one other interpretation is that $\sim$50 per cent of AGN do not have warm absorbers). When observed at high spectral resolution (such as with the transmission gratings on-board {\sl Chandra}), these absorbers are known to produce numerous narrow absorption features ranging from various elements such as oxygen, silicon, neon, carbon, nitrogen, sulphur, magnesium and iron (e.g. Kaspi et al. 2002; Crenshaw, Kraemer \& George 2003; McKernan, Yaqoob \& Reynolds 2007). The associated lines are often blueshifted, thus implying outflowing winds (ranging in velocity from several hundred to several thousand km\,s$^{-1}$; Blustin et al. 2005) and arise from a wide range of column densities and levels of ionisation. Through the study of blueshifted absorption lines of K-shell Fe, recent observations of higher luminosity AGN have also suggested the presence of highly ionised (``hot'') absorbers originating in high-velocity disc winds ($v_{\rm out} \sim -0.1c$) with around a dozen cases to date (e.g. PG 1211+143, Pounds et al. 2003; PDS 456, Reeves, O'Brien \& Ward 2003) although there are a set of lower luminosity sources that still exhibit deep absorption lines at Fe\,K but require high column densities and have outflow velocities on the order of a few thousand km\,s$^{-1}$ (e.g. Mrk 776, Miller et al. 2007; NGC 3516, Turner et al. 2005; NGC 1365, Risaliti et al. 2005; NGC 3783, Reeves et al. 2004). Recently, Tombesi et al. (2010) found evidence for blueshifted features in 17 objects within their sample of 42 radio-quiet AGN observed with {\sl XMM-Newton}, detecting 22 absorption lines at rest-frame energies \textgreater 7.1\,keV. This implies that high velocity outflows may be a common feature in radio-quiet AGN. Such ultra-fast outflows indicate that the gas must be located very close to the nucleus of the AGN and so deep studies of these objects are vital to establish a greater understanding of the outflow kinematics and locations relative to the central continuum source which in turn can help to ultimately unravel the inner structure of AGN. \\

One possible interpretation regarding accretion disc outflows is that they are perhaps produced as a result of radiation pressure a few $R_{\rm g}$ from the event horizon (Proga, Stone \& Kallman 2000). The high luminosities of these systems are the result of radiatively efficient accretion onto a supermassive black hole (SMBH) and such outflows could be a consequence of near-Eddington accretion (King \& Pounds 2003). An alternative interpretation of many outflows is that they could be magnetohydrodynamically (MHD) driven as has been suggested for such objects as GRO J1655-40 (Miller et al. 2008) and NGC 4151 (Kraemer et al. 2005; Crenshaw \& Kraemer 2007). The derived outflow rates can in some cases be comparable to the mass accretion rate (several Solar masses per year) of the AGN and so in terms of kinetic power, the outflow can be responsible for a significant proportion of the bolometric luminosity. As a result, outflows are considered to be an important phenomenon in AGN and are believed to play a key role in feedback processes between the black hole and the host galaxy (King 2003, 2010), ultimately leading to the observed $M$--$\sigma$ relation for galaxies (Ferrarese \& Merritt 2000; Gebhardt et al. 2000). \\

NGC 4051 is a nearby ($z = 0.002336$; Verheijen \& Sancisi 2001) narrow-line Type 1 Seyfert Galaxy at a distance of 15.2\,Mpc (Russel 2004) obtained from the Tully-Fisher relation for nearby galaxies (Tully \& Fisher 1977). It has a black hole mass of $M_{\rm BH} = 1.73^{+0.55}_{-0.52} \times 10^{6}$\,M$_{\odot}$ determined via optical reverberation mapping (Denney et al. 2009) and is well known for its extreme X-ray variability both on short and long time-scales with the X-ray spectrum hardening as the source flux becomes lower (Lamer et al. 2003), as seen in many other Type 1 Seyfert Galaxies (e.g. NGC 5506, Lamer, Uttley \& McHardy 2000; MCG-6-30-15, Vaughan \& Edelson 2001). \\

The warm absorber in NGC 4051 was studied by Collinge et al. (2001) with a simultaneous {\sl Chandra} HETG (High Energy Transmission Grating) and {\sl Hubble Space Telescope} (HST) Imaging Spectrograph observation where they discovered two separate X-ray absorption systems corresponding to outflow velocities of $v_{\rm out} = -(2340 \pm 130)$ and $v_{\rm out} = -(600 \pm 130)$\,km\,s$^{-1}$. Nine separate absorption systems were detected in the HST UV spectrum with one of the zones possibly corresponding to the lower velocity zone detected in the X-ray band. A further RGS (Reflection Grating Spectrometer) observation with {\sl XMM-Newton} was analysed by Ogle et al. (2004) who claimed the presence of a relativistically broadened O\,\textsc{viii} Ly$\alpha$ emission line (EW $\sim 90$\,eV) at $\sim$655\,eV and suggested that this emission along with its associated radiative recombination continuum (RRC) could be responsible for the weak soft excess. However, they found no evidence of the high velocity outflow detected by Collinge et al. (2001). The RGS data from 2001 and 2002 were also analysed by Pounds et al. (2004a) and Nucita et al. (2010). They note that the 2002 observation caught the source during an extended period of low flux ($F_{\rm 2-10} = 5.8 \times 10^{-12}$\,erg\,cm$^{-2}$\,s$^{-1}$). Pounds et al. (2004a) concluded that the hard X-ray spectral shape observed during this period of low flux was due to an increase in opacity of a substantial column of gas in the line-of-sight causing the spectrum to become dominated by a quasi-constant cold reflection component. \\

A more recent study by Krongold et al. (2007) showed the presence of two distinct ionisation components for the absorber through a 100\,ks {\sl XMM-Newton} exposure. By measuring the electron densities through the absorber variability, they inferred that the absorbing components must be compact with the high and low ionisation zones existing at distances of 0.5--1\,l-d (light-days) and \textless 3.5\,l-d from the central engine respectively, well within the dusty torus and strongly suggestive of an accretion disc origin for the warm absorber wind. They calculated that the implied mass outflow rate of the warm absorber wind corresponds to approximately 2--5 per cent of the mass accretion rate of the source. They also detected several narrow emission lines in the RGS spectrum from C\,\textsc{vi}, N\,\textsc{vi}, O\,\textsc{vii}, O\,\textsc{viii}, Ne\,\textsc{ix} and Fe\,\textsc{xvii}. \\

A further study of the warm absorber was also undertaken by Steenbrugge et al. (2009) through high-resolution X-ray spectroscopy using the {\sl Chandra} LETG (Low Energy Transmission Grating). They discovered that this object contains an outflowing wind consisting of four separate absorbing components ranging in ionisation parameter, log\,$\xi$, from 0.07 to 3.19. They found that the absorbing zone with log\,$\xi = 3.19$ requires a high outflow velocity of $v_{\rm out} = -4760$\,km\,s$^{-1}$ and from a study of the variability of the warm absorber, they inferred that three of the four absorbing zones appear to be located in the range of 0.02--1\,pc from the black hole. \\

NGC 4051 was observed in 2005 with {\sl Suzaku} when it was found to have fallen into an extended period of historically low flux (Terashima et al. 2009). A strong excess of emission was seen at energies \textgreater 10\,keV suggesting that the primary power-law continuum had largely disappeared and that the bulk of the observed emission was reflection-dominated (e.g. Guainazzi et al. 1998; Pounds et al. 2004a). However, Terashima et al. (2009) were able to describe the data using a partial covering model whereby the reflection-dominated emission and the intrinsic power-law emission are independently absorbed by gas covering some significant fraction of $4\pi$\,sr. In this scenario, changes in the covering fraction of the partially-covered power-law component can account for the changes in the spectral shape at low energies whereas changes in the normalisation of the power-law component overlaid on a nearly constant hard component can account for the variability at energies \textgreater 3.5\,keV. This interpretation of the long-term spectral variability being caused by the primary X-ray continuum disappearing from view leaving behind a constant, hard, reflection-dominated component was supported by the results of Miller et al. (2010) through principal components analysis (PCA) of the 2005 and 2008 {\sl Suzaku} data. Miller et al. (2010) also suggested that the low-flux states are caused by variable partial covering obscuring the central engine which also explains the constancy of the narrow Fe\,K$\alpha$ emission line flux. \\

Finally, the presence of an additional absorption feature at $\sim$7.1\,keV was reported by Pounds et al. (2004a) indicating that a highly ionised, high velocity outflow could be apparent in this source. Further observations with {\sl Suzaku} and {\sl XMM-Newton} have confirmed the presence of this feature along with an additional absorption line at $\sim$6.8\,keV (Terashima et al. 2009; Pounds \& Vaughan 2010, submitted). If the two absorption lines are associated with K-shell transitions from Fe\,\textsc{xxv} and Fe\,\textsc{xxvi} then their blueshift would imply that the material is outflowing with a velocity of $v_{\rm out} \sim -0.02c$. \\

Here we report on a $\sim$300\,ks {\sl Chandra} HETG (High Energy Transmission Grating) observation of NGC 4051 where we aim to study the soft X-ray warm absorber in detail (Sections 4 and 5). We also report on a contemporaneous 350\,ks {\sl Suzaku} observation where we study the Fe\,K band including the blueshifted Fe\,K absorption lines as the signature of the highest ionisation component of the outflowing wind. We then proceed to analyse the entire broad-band {\sl Suzaku} spectrum from 0.5--50.0\,keV. Finally, with a combined analysis of the 2005 archival {\sl Suzaku} data (exposure time $\sim$ 120\,ks) when the source was found to be in an extended period of low flux, we also aim to study the origin of the long-term X-ray variability of this AGN (Section 6).

\section{Observations and Data Reduction}

\subsection{Chandra Analysis}

NGC 4051 was observed with the {\sl Chandra} X-ray Observatory (Weisskopf et al. 2000) on twelve different occasions dating from 2008 Nov 6 to 2008 Nov 30 with a total exposure time of $\sim$300\,ks. A log of the separate observations and their corresponding exposures is shown in Table 1. The {\sl Chandra} observations were performed with the High Energy Transmission Grating Spectrometer (HETGS; Markert et al. 1994; Canizares et al. 2005) in the focal plane of the Advanced CCD Imaging Spectrometer (ACIS; Garmire et al. 2003) which consists of two separate gratings: the Medium Energy Grating (MEG; 0.4--8.0\,keV) and the High Energy Grating (HEG; 0.7--10.0\,keV). The {\sl Chandra} HETG data were reduced using version 4.1 of both the Chandra Interactive Analysis of Observations software package (\textsc{ciao}\footnote{http://cxc.harvard.edu/ciao}; Fruscione et al. 2006) and corresponding Calibration Database (CALDB).

\subsubsection{HETG Reduction}

First order MEG and HEG spectra were extracted for the source and background for each individual observation. Spectral redistribution matrix (RMF) files were created using the \textsc{mkgrmf} script for each first order grating arm (-1 and +1) for the MEG and HEG. Telescope effective area files were also created using the \textsc{fullgarf} script incorporating the \textsc{ciao} tool \textsc{mkgarf}. Events files were extracted from the negative and positive first order grating arms of the MEG and HEG allowing light curves and spectra to be extracted. The first order spectra were then combined for each individual exposure using combined response files (with appropriate weighting) for the MEG and HEG. The background was not subtracted however as it has a negligible effect in the energy ranges of interest here. The total count rates in the first-order energy spectra (0.5--10.0\,keV) are $0.602\pm0.001$ and $0.250\pm0.001$\,ct\,s$^{-1}$ for the MEG and HEG respectively (or $0.583\pm0.001$ and $0.233\pm0.001$\,ct\,s$^{-1}$ for the respective 0.5--5.0 and 1.0--8.0\,keV energy bands for the MEG and HEG which we adopt here). These count rates correspond to fluxes of $3.99 \times 10^{-11}$ and $3.29 \times 10^{-11}$\,erg\,cm$^{-2}$\,s$^{-1}$ (see Table 2) and luminosities of $4.78 \times 10^{41}$ and $3.95 \times 10^{41}$\,erg\,s$^{-1}$ for the respective 0.5--5.0 and 1.0--8.0\,keV energy bands for the MEG and HEG. A lightcurve showing the twelve individual observations with 2\,ks binning over the 1.0--6.0\,keV range (including both the MEG and HEG) is shown in Figure 1.

\begin{figure}
\begin{center}
\rotatebox{-90}{\includegraphics[width=6cm]{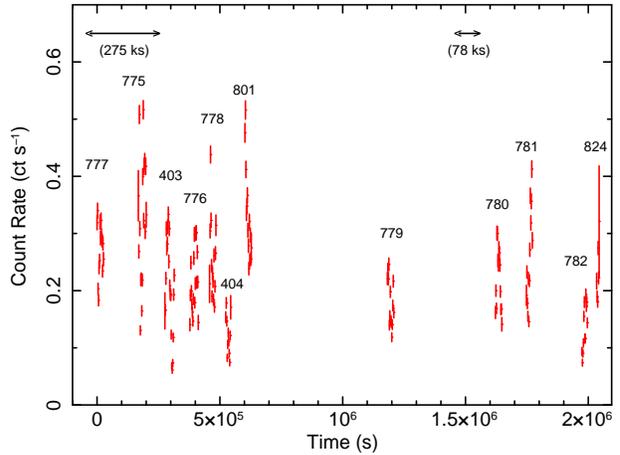}}
\end{center}
\caption{The summed HEG and MEG lightcurve showing the count rates of the twelve individual {\sl Chandra} HETG observations of NGC 4051 over the 1.0--6.0\,keV energy range. The labels refer to the sequence numbers of the observations listed in Table 1. The data are binned up in 2\,ks bins and the horizontal lines show the duration and overlap of the two contemporaneous {\sl Suzaku} observations.} 
\end{figure}

\subsection{Suzaku Analysis}

NGC 4051 was observed with {\sl Suzaku} in November 2005 and again in November 2008 with net exposures of 120 and 350\,ks respectively\footnote{Note that the 2008 {\sl Suzaku} observation was split into two separate observations due to scheduling constraints with respective exposure times of 275 and 78\,ks.} (see Table 1 for an observation log). The 2005 data have been described previously by Terashima et al. (2009). Here we discuss data taken with the {\it Suzaku} XIS (X-ray Imaging Spectrometer; Koyama et al. 2007), consisting of four X-ray telescopes (Mitsuda et al. 2007) each with a CCD in the focal plane, and the PIN diodes of the non-imaging HXD (Hard X-ray Detector; Takahashi et al. 2007). Note that NGC 4051 is too faint to be detected with the HXD GSO instrument. Events files from versions 1.2 and 2.2.11.22 of the {\sl Suzaku} pipeline processing were used for the 2005 and 2008 observations respectively. All data were reduced using HEA\textsc{soft} version 6.4.1.

\subsubsection{XIS Reduction}

A full description of the {\sl Suzaku} XIS data reduction procedure can be found in Turner et al. (2010) since the same base files are used here. The count rates corresponding to the 2005 Nov 10, 2008 Nov 6 and 2008 Nov 23 observations respectively over the 0.5--10.0\,keV energy range were found to be $\sim$0.45, $\sim$2.07 and $\sim$1.42\,ct\,s$^{-1}$ per front-illuminated XIS with the background rates corresponding to $\sim$2.4 per cent, $\sim$0.9 per cent and $\sim$0.9 per cent of the respective source count rates. These count rates correspond to fluxes of $1.32 \times 10^{-11}$, $5.03 \times 10^{-11}$ and $3.53 \times 10^{-11}$\,erg\,cm$^{-2}$\,s$^{-1}$ and luminosites of $3.66 \times 10^{41}$, $1.57 \times 10^{42}$ and $1.10 \times 10^{42}$\,erg\,s$^{-1}$ for the three respective observations. The three front-illuminated XIS chips (XIS 0,\,2,\,3; hereafter XIS--FI) were used in the spectral analysis of the 2005 data due to their greater sensitivity at Fe\,K. However, as the use of XIS 2 was discontinued after a charge leak was discovered in 2006, the spectral analysis from the 2008 data refers only to XIS 0 and 3. The XIS 1 (back-illuminated) data were checked for consistency but not used during the spectral fitting due to the detector's reduced sensitivity compared to XIS--FI at Fe\,K. As the chips were found to produce consistent spectra within the statistical errors in all observations, the XIS--FI spectra and responses were combined to maximise signal to noise. The XIS source spectra were binned at the half-width at half-maximum (HWHM) resolution of the detector due to the high photon statistics. This enabled the use of $\chi^{2}$ minimisation in all {\sl Suzaku} fits as there were \textgreater 50 counts per resolution bin.

\subsubsection{HXD Reduction}

Likewise, the {\sl Suzaku} HXD data reduction procedure can also be found in Turner et al. (2010). As NGC 4051 is too faint to be detected by the HXD GSO instrument, we use data taken with the HXD PIN only which provides useful data over the 15.0--70.0\,keV energy range. Our net exposure times of the PIN source spectra were found to be 112, 204 and 59\,ks for the 2005 Nov 10, 2008 Nov 6 and 2008 Nov 23 observations respectively. Note that the HXD PIN response file dated 2008/01/29 (epoch 1) for the HXD nominal pointing position was used for the 2005 observation whereas the response file dated 2008/07/16 (epoch 5) for the XIS nominal pointing position was used for the the 2008 observations. For the three respective observations, the net PIN source count rates are $0.043\pm0.001$, $0.068\pm0.001$ and $0.053\pm0.001$\,ct\,s$^{-1}$ over the 15--50\,keV range. These count rates in turn correspond to fluxes of $1.73 \times 10^{-11}$, $3.06 \times 10^{-11}$ and $2.66 \times 10^{-11}$\,erg\,cm$^{-2}$\,s$^{-1}$ for the three respective observations. The mean net count rates and fluxes are summarised in Table 2.

\begin{table}
\centering
\caption{{\sl Suzaku} and {\sl Chandra} HETG observation log of NGC 4051.}
\begin{tabular}{l c c c}
\hline\hline
\multirow{2}{*}{Mission} & \multirow{2}{*}{Sequence Number} & Observation Date & \multirow{2}{*}{Exposure (ks)} \\
& & (Start Date) & \\
\hline
{\sl Suzaku} & 700004010 & 2005-11-10 & 120 \\
{\sl Suzaku} & 703023010 & 2008-11-06 & 275 \\
{\sl Chandra} & 10777 & 2008-11-06 & 27.8 \\
{\sl Chandra} & 10775 & 2008-11-08 & 30.9 \\
{\sl Chandra} & 10403 & 2008-11-09 & 38.2 \\
{\sl Chandra} & 10778 & 2008-11-11 & 34.1 \\
{\sl Chandra} & 10776 & 2008-11-11 & 25.2 \\
{\sl Chandra} & 10404 & 2008-11-12 & 20.1 \\
{\sl Chandra} & 10801 & 2008-11-13 & 26.1 \\
{\sl Chandra} & 10779 & 2008-11-20 & 27.8 \\
{\sl Suzaku} & 703023020 & 2008-11-23 & 78 \\
{\sl Chandra} & 10780 & 2008-11-25 & 26.4 \\
{\sl Chandra} & 10781 & 2008-11-26 & 24.1 \\
{\sl Chandra} & 10782 & 2008-11-29 & 23.7 \\
{\sl Chandra} & 10824 & 2008-11-30 & 9.2 \\
\hline
\end{tabular}
\end{table}

\begin{table*}
\centering
\caption{Mean count rates and fluxes for the time-averaged {\sl Chandra} HETG and {\sl Suzaku} observations of NGC 4051. All XIS, HXD, MEG and HEG count rates and fluxes are given in the 0.5--10.0, 15.0--50.0, 0.5--5.0 and 1.0--8.0\,keV energy bands respectively.}
\begin{tabular}{l c c c c}
\hline\hline
Mission & Instrument & Observation Date & Count Rate & Flux \\ [0.5ex]
& & & (ct\,s$^{-1}$) & ($\times 10^{-11}$\,erg\,cm$^{-2}$\,s$^{-1}$) \\
\hline
\multirow{2}{*}{{\sl Suzaku}} & XIS & \multirow{2}{*}{2005-11-10} & 0.45 & 1.32 \\
& HXD & & 0.04 & 1.73 \\
\multirow{2}{*}{{\sl Suzaku}} & XIS & \multirow{2}{*}{2008-11-06} & 2.07 & 5.03 \\
& HXD & & 0.07 & 3.06 \\
\multirow{2}{*}{{\sl Chandra}} & MEG & \multirow{2}{*}{2008-11-06} & 0.58 & 3.99 \\
& HEG & & 0.23 & 3.29 \\
\multirow{2}{*}{{\sl Suzaku}} & XIS & \multirow{2}{*}{2008-11-23} & 1.42 & 3.53 \\
& HXD & & 0.05 & 2.66 \\
\hline
\end{tabular}
\end{table*}

\section{Spectral Analysis}

For all spectral analysis we used the \textsc{xspec v11.3} software package (Arnaud 1996). All our fits include Galactic absorption with a column density $N^{\rm GAL}_{\rm H} = 1.35\times10^{20}$\,cm$^{-2}$, obtained using the \textsc{ftool nh} with the compilations of Dickey \& Lockman (1990). We also used the cross-sections for X-ray absorption by the interstellar medium obtained by Morrison \& McCammon (1983) by using \textsc{wabs} in \textsc{xspec}. Abundances are those of Anders \& Grevesse (1989) unless otherwise stated. Due to the low number of counts per channel (i.e. \textless 50 counts per resolution bin) in the HETG data, we could not use $\chi^{2}$ minimisation; instead, we checked our fit statistics in the {\sl Chandra} spectra using the $C$-statistic (Cash 1979). However, due to the higher photon statistics in the {\sl Suzaku} spectra, all goodness-of-fit values were checked using $\chi^{2}$ minimisation. Note that all errors quoted correspond to 90 per cent confidence for one interesting parameter ($\Delta C = \Delta \chi^{2} = 2.71$) unless stated otherwise. Where $\Delta C$ or $\Delta \chi^{2}$ values are quoted in tables, these values have been determined by removing the component from the final model and re-fitting. All fit parameters are given in the rest frame of the host galaxy having been corrected for the cosmological redshift ($z = 0.002336$) and assuming a distance of 15.2\,Mpc to NGC 4051 (Russell 2004) obtained from the Tully-Fisher relation for nearby galaxies (Tully \& Fisher 1977).

\section{The Chandra HETG Spectrum At Low Energies}

We initially considered the twelve separate HETG observations (see Table 1) and found that the flux appears to vary by a factor of $\sim$2 over the course of this 24-day period. The short-term variability appears to be consistent with bright narrow-line Type 1 Seyfert Galaxies in that the X-ray continuum softens as the flux increases (e.g. Lamer et al. 2003; Taylor, Uttley \& McHardy 2003). However, we find that the low signal-to-noise ratio due to these short individual exposures is insufficient to significantly determine the parameters of the warm absorber and detect individual absorption lines so we begin by considering the entire 300\,ks time-averaged HETG spectrum. We initially binned the MEG and HEG data to a constant resolution of $\Delta \lambda = 0.01$\,\AA\ and $\Delta \lambda = 0.005$\,\AA\ respectively and in all subsequent fits we consider the MEG and HEG over the respective 0.5--5.0\,keV and 1.0--8.0\,keV energy bands.

\subsection{Continuum}

We initially fitted a simple parameterisation of the continuum using the MEG and HEG in order to allow the warm absorber to be studied in detail. The continuum emission was fitted over the 0.5--8.0\,keV range with an absorbed power-law with a Galactic column density $N^{\rm GAL}_{\rm H} = 1.35\times10^{20}$\,cm$^{-2}$ (Dickey \& Lockman 1990) and a photon index $\Gamma = 2.12\pm0.01$. This gave a poor fit to the data with $C / d.o.f. = 6082 / 4396$. The fluxed HETG spectrum unfolded against a power-law with $\Gamma = 2$ is shown in Figure 2 where significant emission can be observed \textgreater 6\,keV due to the Fe\,K emission complex and positive spectral curvature can also be seen below 1\,keV indicating that an additional steepening of the spectrum is required in this source. We parameterised this with the addition of a simple featureless black body ($kT = 0.10\pm0.01$\,keV) which improved the fit by $\Delta C = 197$. The purpose of the black body component was not to model the data in a physical sense but to simply parameterise the soft excess with a smooth continuum, therefore allowing the individual lines from the warm absorber to be studied in detail. This yields a decent fit to the low-energy continuum and as we see no evidence for any emission / absorption features \textgreater 2\,keV (apart from at Fe\,K), we excluded the MEG and HEG data \textgreater 2\,keV and proceeded to analyse and identify the soft ionised absorption lines in the 0.5--2.0\,keV band. A plot of the $\sigma$ residuals \textless 2\,keV is shown in Figure 3 revealing a number of absorption and emission lines.

\begin{figure}
\begin{center}
\rotatebox{-90}{\includegraphics[width=6cm]{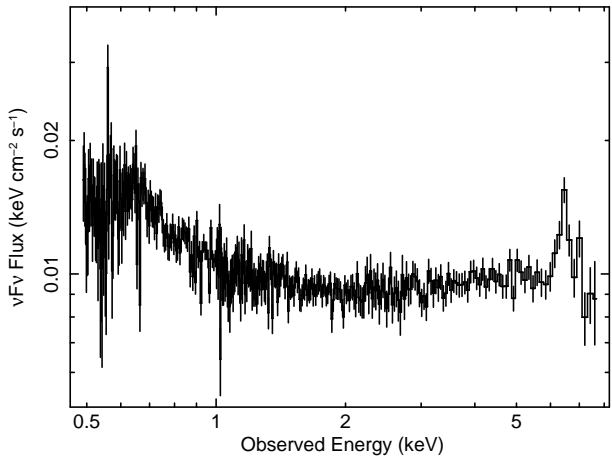}}
\end{center}
\caption{A plot of the HETG spectrum from 0.5--8.0\,keV unfolded against a power-law with $\Gamma = 2$. Strong emission at energies \textgreater 6\,keV is most likely due to significant Fe\,K emission and positive spectral curvature can also be observed at energies \textless 1\,keV suggestive of the presence of a weak soft excess. The MEG and HEG are fitted over the 0.5--5.0 and 1.0--8.0\,keV bands respectively (in the observed frame) and are binned by a factor of 10 for clarity.} 
\end{figure}

\begin{figure*}
\begin{center}
\rotatebox{-90}{\includegraphics[height=15.3cm]{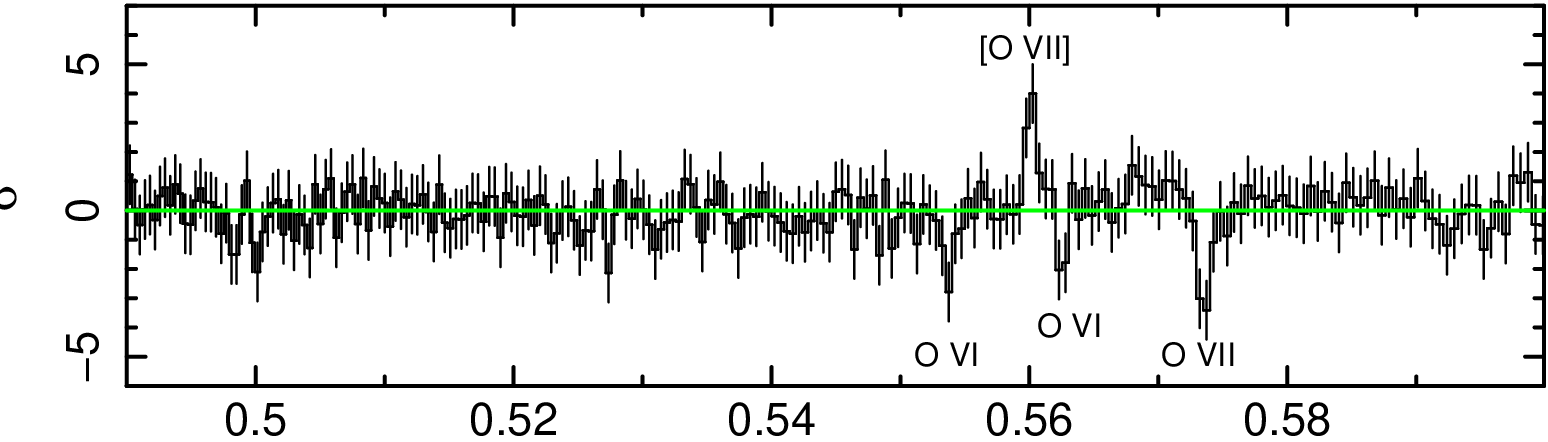}}
\rotatebox{-90}{\includegraphics[height=15.3cm]{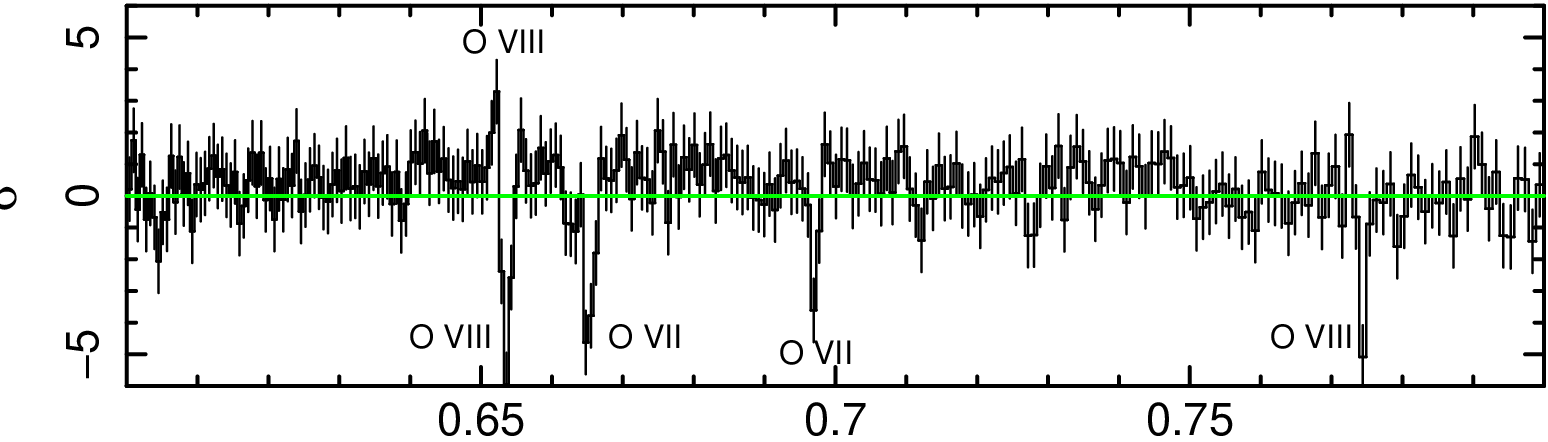}}
\rotatebox{-90}{\includegraphics[height=15.3cm]{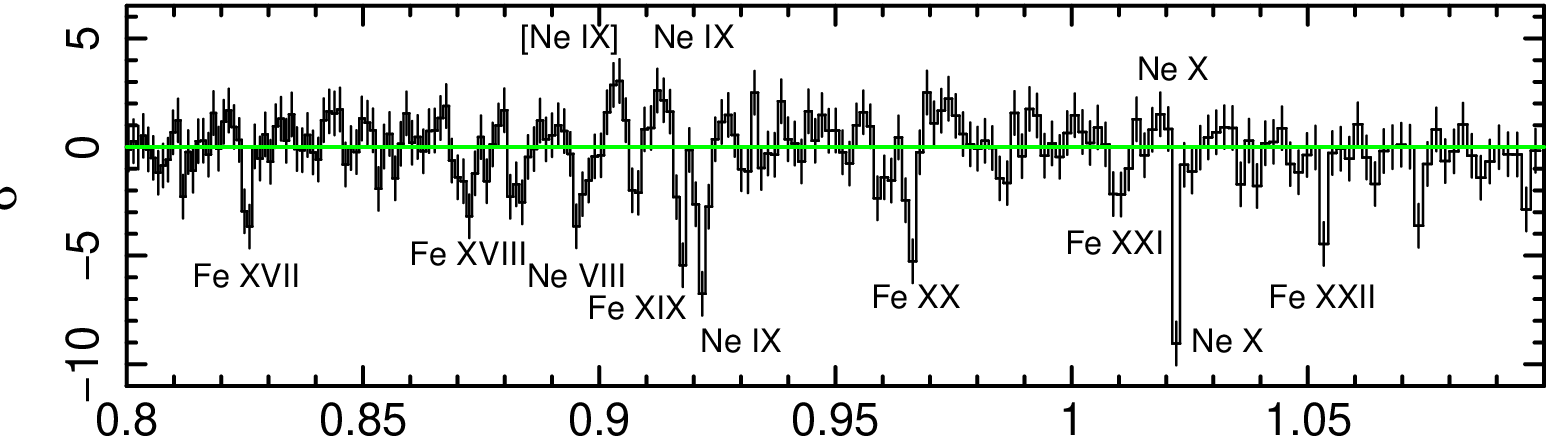}}
\rotatebox{-90}{\includegraphics[height=15.3cm]{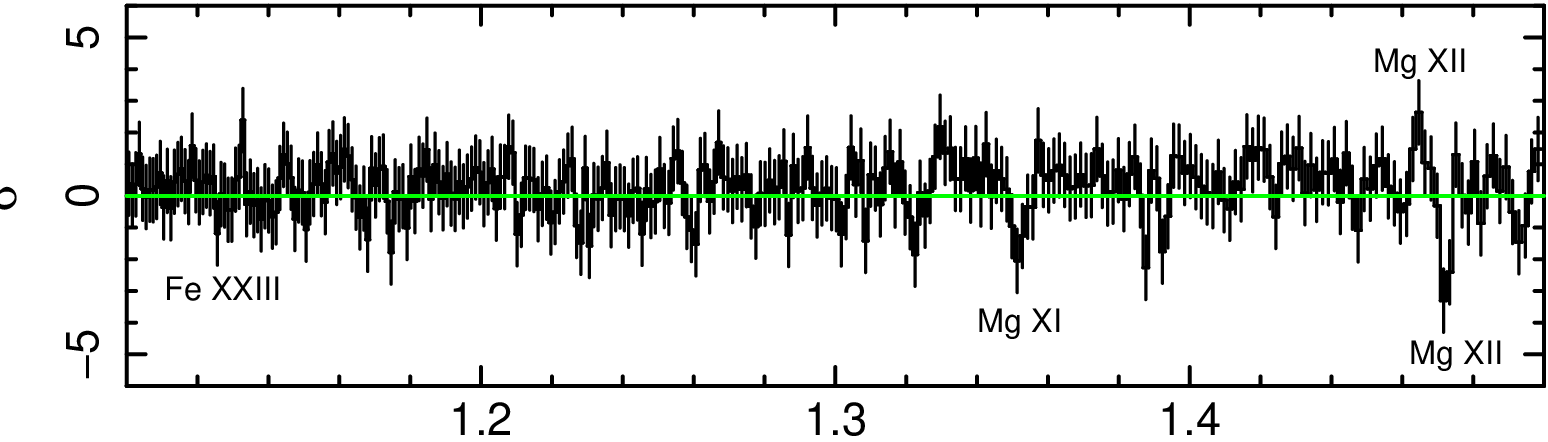}}
\rotatebox{-90}{\includegraphics[height=15.3cm]{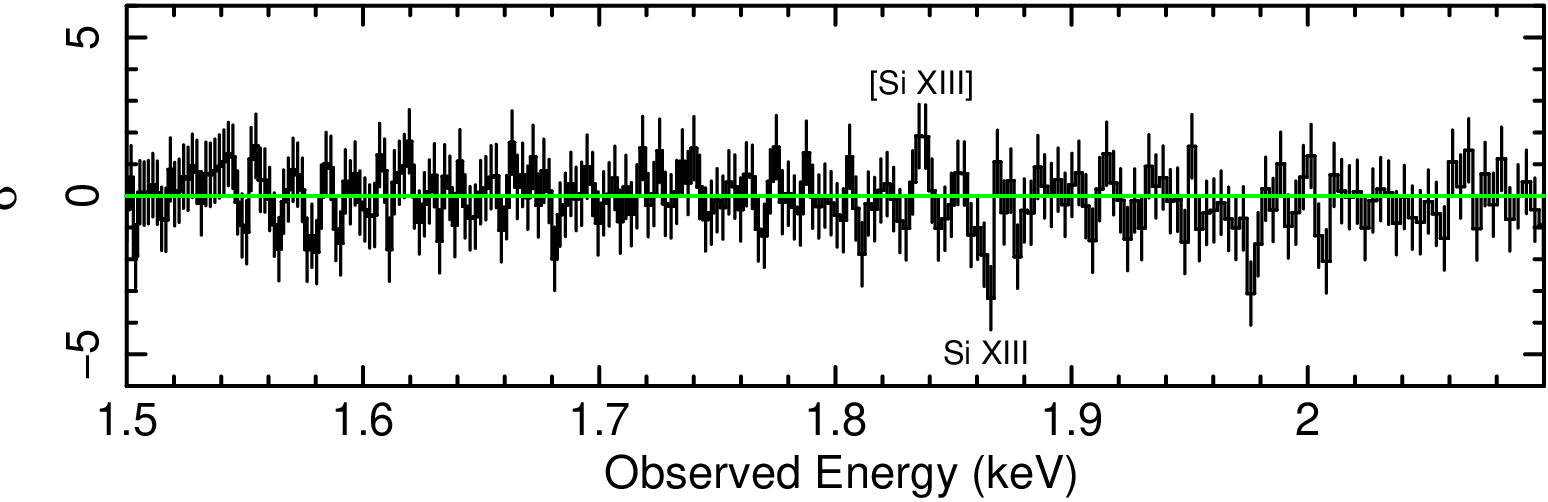}}
\end{center}
\caption{The $\sigma$ residuals of the HETG spectrum from 0.5--2.0\,keV when fitted with a single absorbed power-law and a featureless blackbody as described in Section 3.1. Significant residuals can be observed throughout the spectrum indicating the presence of several absorption and emission features. The statistically significant features which we are able to identify are labelled on the plot. All spectra are binned up by a factor of 2 for clarity and are plotted in the observed frame.} 
\end{figure*}

\subsection{Absorption Lines}

The time-averaged 300\,ks spectrum reveals 21 significant (\textgreater 99.9 per cent) soft X-ray absorption lines below 2.0\,keV originating from material with a range of ionisation states and column densities. Several of the lines appear to be somewhat resolved in the HETG spectrum (e.g. O\,\textsc{vii} $1s$--$2p$, O\,\textsc{viii} $1s$--$2p$, Fe\,\textsc{xviii} $2p$--$3d$) with FWHM on the order of a few hundred to $\sim$1\,000\,km\,s$^{-1}$. Indeed we find a mean value for the FWHM of $580 \pm 59$\,km\,s$^{-1}$ with a dispersion of $\sigma = 210$\,km\,s$^{-1}$ for the 13 absorption lines which appear to be resolved in the spectrum.  
The spectrum reveals absorption primarily from H-like and He-like ions of oxygen, neon, magnesium and silicon along with lines from less-ionised ions such as O\,\textsc{vi}. We also detect several L-shell transitions of Fe\,\textsc{xvii}-Fe\,\textsc{xxii}. After taking into account the systemic redshift of NGC 4051, we note that the observed centroid energies of the absorption lines are all blueshifted on the order of a couple of eV. This blueshift implies that the absorbing material in our line-of-sight is outflowing with calculated velocity shifts on the order of a few hundred to $\sim$1\,000\,km\,s$^{-1}$ (e.g. the O\,\textsc{vii}, O\,\textsc{viii} $1s$--$2p$ lines) relative to the host galaxy. A mean value for the outflow velocity of $v_{\rm out} = -(620 \pm 34)$\,km\,s$^{-1}$ with a dispersion of $\sigma = 150$\,km\,s$^{-1}$ is calculated for the 20 absorption lines that we are able to identify. We determined the parameters of the lines by modelling them with simple Gaussian profiles. These parameters and their likely identifications are summarised in Table A1. All lines were detected at the \textgreater 99.9 per cent significance level for two interesting parameters (i.e. $\Delta C \geq 13.8$).

\subsection{Emission Lines}

In addition to a wealth of absorption features, the HETG spectrum also reveals 7 narrow emission lines which were again initially modelled with simple symmetric Gaussians. Their identifications appear to be largely consistent with the narrow emission lines found in the {\sl XMM-Newton} RGS spectrum of NGC 4051 by Pounds et al. (2004a) and also with those detected by Terashima et al. (2009). These lines appear to primarily originate from He-like ions of oxygen, neon and silicon due to their associated forbidden transitions. The Ne\,\textsc{ix} intercombination lines are also identified at $\sim$915\,eV\footnote{Interestingly, the forbidden and intercombination lines from Ne\,\textsc{ix} appear to have similar strengths; this could be due to the presence of the Ne\,\textsc{viii} $1s$--$2p$ absorption line at $\sim$895\,eV reducing the observed flux of the Ne\,\textsc{ix} forbidden transition, thus making it appear to be weaker than it really is.}. Furthermore, Ly$\alpha$ transitions from H-like species of oxygen and magnesium are also observed at $\sim$653\,eV and $\sim$1\,472\,eV respectively. We note that the O\,\textsc{viii} Ly$\alpha$ emission that we observe here appears to be narrow (FWHM \textless 400\,km\,s$^{-1}$; see Figure 4) and that we do not require an additional relativistically broadened Ly$\alpha$ feature such as the one claimed by Ogle et al. (2004). The emission lines that we detect here have a mean velocity shift, $v_{\rm out}$, within $\pm 200$\,km\,s$^{-1}$ of the systemic shift and appear to be unresolved except for the forbidden emission lines from Ne\,\textsc{ix} (FWHM $= 550^{+370}_{-500}$\,km\,s$^{-1}$) and Si\,\textsc{xiii} (FWHM $= 1\,000^{+790}_{-450}$\,km\,s$^{-1}$). We do note that the Ly$\alpha$ transitions of O\,\textsc{viii} and Mg\,\textsc{xii} appear to be redshifted on the order of a few 100\,km\,s$^{-1}$ relative to their rest frame centroid energies. This is most likely due to the P-Cygni-like profiles in which the lines appear with their corresponding absorption lines (see Figure 4). The best-fitting parameters of the emission lines are listed in Table A1. Like the absorption features, all of the emission lines were also detected at the \textgreater 99.9 per cent significance level for two interesting parameters (i.e. $\Delta C \geq 13.8$).

\section{The Broad-band HETG Spectrum}

In order to model the broad-band HETG spectrum we included the data above 2\,keV in our spectral fits so that we were considering the full 0.5--8.0\,keV energy range. We carried forward the model from Section 3 consisting of a simple baseline continuum, a featureless blackbody to parameterise the soft spectral steepening and symmetric Gaussians to model the soft absorption and emission features. Figure 2 shows that a significant excess can be observed in the Fe\,K band at energies \textgreater 6\,keV (see Section 5.2). However, we begin by considering a more physical parameterisation of the warm absorber. \\

\subsection{The Warm Absorber}

We removed the simple Gaussian profiles parameterising the absorption lines from our model and instead attempted to model the warm absorber using the \textsc{xstar} 2.1ln11 code of Kallman \& Bautista (2001; also see Kallman et al. 2004), incorporating the abundances of Grevesse, Noels \& Sauval (1996). The \textsc{xstar} code assumes thin shells of absorbing gas and self-consistently models zones of absorption parameterised by their column density, $N_{\rm H}$, and ionisation parameter, $\xi$, which is defined as:

\begin{equation}\xi = \frac{L_{\rm ion}}{nR^{2}}\end{equation}

and has units erg\,cm\,s$^{-1}$, where $L_{\rm ion}$ is the ionising luminosity from 1 to 1\,000 Rydbergs in units erg\,s$^{-1}$, $n$ is the gas density in cm$^{-3}$ and $R$ is the radial distance in cm of the absorbing gas from the central source of X-rays. For the spectral energy distribution (SED) of the \textsc{xstar} models we assume a simple illuminating power law with $\Gamma = 2.5$, although this is difficult to compare with the observed SED of NGC 4051 since the UV data from the HST were acquired a year later and so are not simultaneous. We initially attempted to model the absorption lines using Solar abundances and a turbulent velocity width of $\sigma = 200$\,km\,s$^{-1}$, a value largely comparable to the observed velocity widths from the Gaussian line fits (see Table A1). \\

We find that we statistically require four individual zones of absorbing gas to model the data covering four distinct levels of ionisation. Each zone is significant at the \textgreater 99.9 per cent confidence level. The column densities are found to be on the order of $N_{\rm H} \sim 10^{20}$--$10^{21}$\,cm$^{-2}$ and the zones appear to cover a range in ionisation parameter of log\,$\xi = -0.86$ to log\,$\xi = 2.97$. All the zones appear to be blueshifted implying that the material is outflowing with velocities on the order of a few 100\,km\,s$^{-1}$ with the general trend being that $v_{\rm out}$ increases with increasing $\xi$. We find that one of the zones requires two significantly different outflow velocities and so an additional zone with comparable values for the ionisation parameter and column density is included in the fit to account for this (see zones 3a and 3b in Table 3). We note that replacing zones 3a and 3b with one zone of higher turbulence velocity significantly worsens the fit. Therefore, the inclusion of the two separate zones with $\sigma = 200$\,km\,s$^{-1}$ and differing outflow velocities appears to be preferred by the data. Note that we subsequently refer to zones 3a and 3b as one single zone. A summary of the best-fitting values and the corresponding fit statistic ($\Delta C$) for each zone is given in Table 3. \\

Regarding the transitions modelled by each \textsc{xstar} zone, we find that the first zone (ultra-low ionisation) is mainly responsible for the absorption of O\,\textsc{vi} and lower ionisation O ions. The second zone, with a slightly higher ionisation parameter, appears to model absorption primarily from O\,\textsc{vii}. An inspection of the residuals reveals that this zone under-predicts the amount of absorption at $\sim$666\,eV which is likely due to the turbulence velocity of $\sigma = 200$\,km\,s$^{-1}$ being too low. This has the effect of saturating the O\,\textsc{vii} $1s$--$3p$ line and hence underpredicting its EW. Increasing the turbulence velocity of this zone to $\sigma = 500$\,km\,s$^{-1}$ significantly improves the fit and better models the residuals at $\sim$666\,eV (see Figure 4). However, we find that increasing the turbulence velocity of the remaining warm absorber zones to $\sigma = 500$\,km\,s$^{-1}$ significantly worsens the fit and that these zones are better modelled with $\sigma = 200$\,km\,s$^{-1}$. Zones 3a and 3b appear to adequately model the absorption from ions such as Ne\,\textsc{ix}, O\,\textsc{viii} and the various L-shell transitions from Fe\,\textsc{xvii - xxii} and finally, the highest ionisation zone appears to correspond to absorption from ions such as Ne\,\textsc{x}, Mg\,\textsc{xi - xii} and Si\,\textsc{xiii}. A plot of the warm absorber model superimposed on the data is shown in Figure 5 and a further plot showing the individual contribution of each of the four zones is shown in Figure 6. A subsequent paper (Crenshaw et al. 2010 in preparation) will discuss the results of a UV observation of NGC 4051 taken with the Cosmic Origins Spectrograph (COS) on-board the HST. The preliminary results suggest that the X-ray zones 1 and 2 detected here with {\sl Chandra} may also coincide with absorption in the UV data. \\

\begin{figure}
\begin{center}
\rotatebox{-90}{\includegraphics[width=6cm]{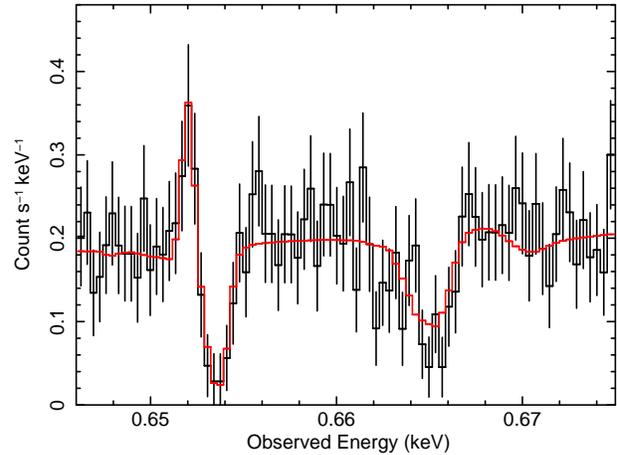}}
\end{center}
\caption{A plot showing the relatively narrow O\,\textsc{viii} Ly$\alpha$ emission at $\sim$652\,eV in the observed frame. This line, along with its corresponding absorption feature at $\sim$653\,eV appears to be P-Cygni-like in appearance. 
The model is superimposed on the data and is shown in red; the O\,\textsc{viii} Ly$\alpha$ emission at $\sim$652\,eV is modelled with a Gaussian and the corresponding absorption line at $\sim$653\,eV is modelled with \textsc{xstar} as described in Section 5.1. We note that a turbulence velocity of $\sigma = 500$\,km\,s$^{-1}$ in the warm absorber model (Zone 2) is required to model the absorption line at $\sim$666\,eV which likely corresponds to the $1s$--$3p$ transition from O\,\textsc{vii}. } 
\end{figure}

We note there are no detected absorption lines associated with a fifth warm absorber zone. However, we did allow for the presence of a partially-covering zone of absorption by including an additional power-law component absorbed by a further shell of gas. We tied the photon index of the two power-law components together leaving just one of the components absorbed by an \textsc{xstar} zone. We note that zones 1--4 of the warm absorber and a column of Galactic hydrogen remained in the model absorbing the entire continuum. We find that this additional partial-coverer improves the fit significantly with $\Delta C = 286$ and appears to be important in modelling additional curvature in the continuum at energies below about 5\,keV. The best-fit parameters of the partial-coverer indicate that the material is partially-ionised with $\log \xi < 2.40$ and has a substantial line-of-sight column density, $N_{\rm H} = 2.00^{+0.68}_{-0.53} \times 10^{23}$\,cm$^{-2}$. The addition of this component also allows the temperature of the black body component to re-adjust slightly to $kT = 0.12 \pm 0.01$\,keV (a value consistent with Terashima et al. 2009) and the best-fitting photon index of the power-law continuum to steepen to $\Gamma = 2.25^{+0.02}_{-0.01}$. Indeed, without the partial-coverer, we note that the photon index of $\Gamma \sim 2.1$ appears to be slightly too flat for this source in the high-flux state (e.g. Lamer et al. 2003; Miller et al. 2010). The normalisation ratio of the absorbed to total power-law components at 1\,keV is $4.98 \times 10^{-3} / 1.67 \times 10^{-2} \approx 0.30$ which suggests that the partial-coverer may correspond to a line-of-sight covering fraction of $\sim$30 per cent of the nuclear X-ray source. The best-fitting parameters of the partial-coverer are also noted in Table 3. \\             

Furthermore, replacing the black body with the \textsc{reflionx} reflection model of Ross \& Fabian (2005) provides an equally good parameterisation of the soft excess. We find that no relativistic blurring is required and that the addition of this component allows the photon index of the power-law to steepen further with $\Gamma = 2.45 \pm 0.02$, a value consistent with that found from a quasi-simultaneous broad-band {\sl Suzaku} observation in 2008 (see Section 6 where this is discussed in more detail). The \textsc{reflionx} component has a best-fitting value for the ionisation parameter of log\,$\xi = 3.3 \pm 0.1$ and corresponds to a ratio of reflected flux to incident flux of $\sim$0.8 over the 0.5--8.0\,keV energy range. However, the ionised reflector is unable to account for some of the soft X-ray emission lines, particularly the forbidden transitions from O\,\textsc{vii}, Ne\,\textsc{ix} and Si\,\textsc{xiii} which indicates that these lines could have an alternative origin. There have been previous claims in the literature of detections of RRC associated with H-like and He-like ions of C, N, O and Ne (e.g. Ogle et al. 2004; Pounds et al. 2004a) using the RGS on-board {\sl XMM-Newton} which suggests that the soft X-ray emission lines could instead be the signature of a photo-ionised plasma. Indeed it has been somewhat well established that such soft X-ray lines are commonly formed either in the narrow-line region (NLR) or as part of the photo-ionised outflow (e.g. Bianchi, Guainazzi \& Chiaberge 2006).

\begin{figure*}
\begin{center}
\rotatebox{-90}{\includegraphics[height=15.3cm]{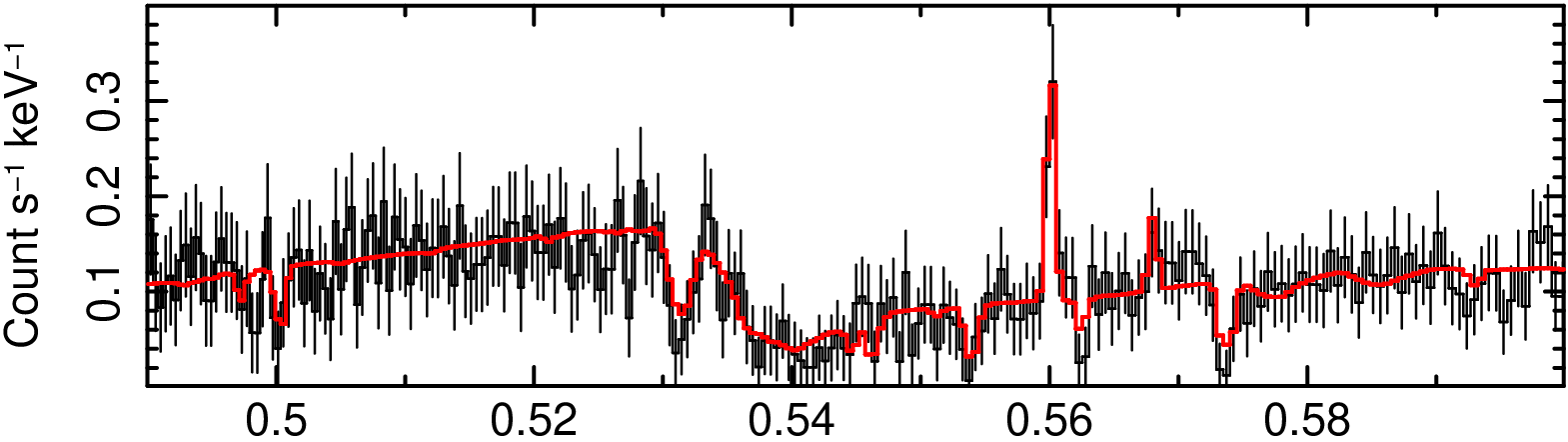}}
\rotatebox{-90}{\includegraphics[height=15.3cm]{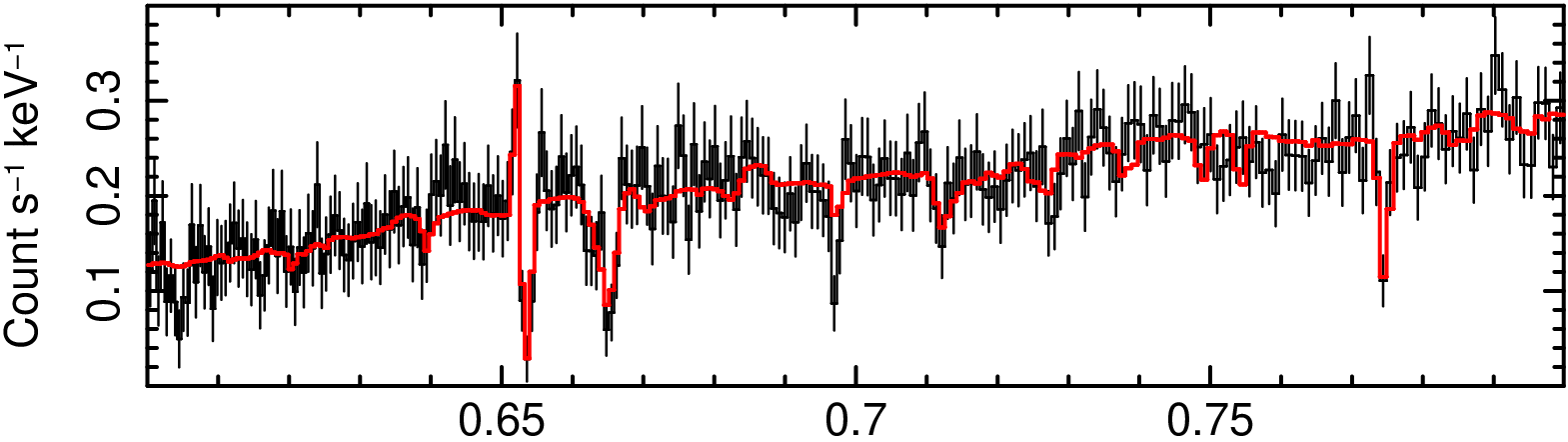}}
\rotatebox{-90}{\includegraphics[height=15.3cm]{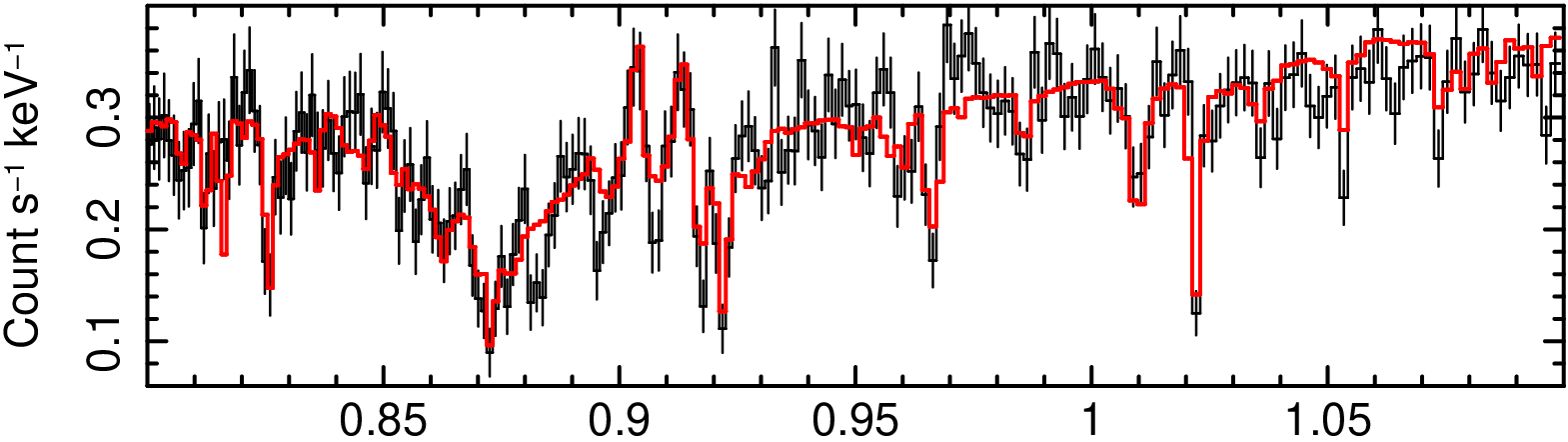}}
\rotatebox{-90}{\includegraphics[height=15.3cm]{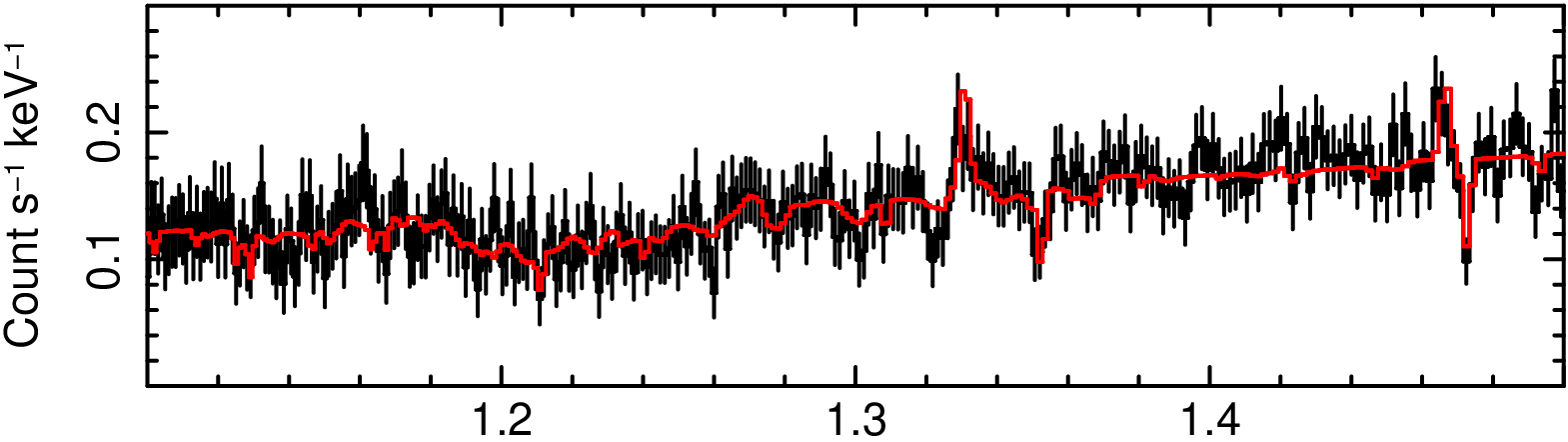}}
\rotatebox{-90}{\includegraphics[height=15.3cm]{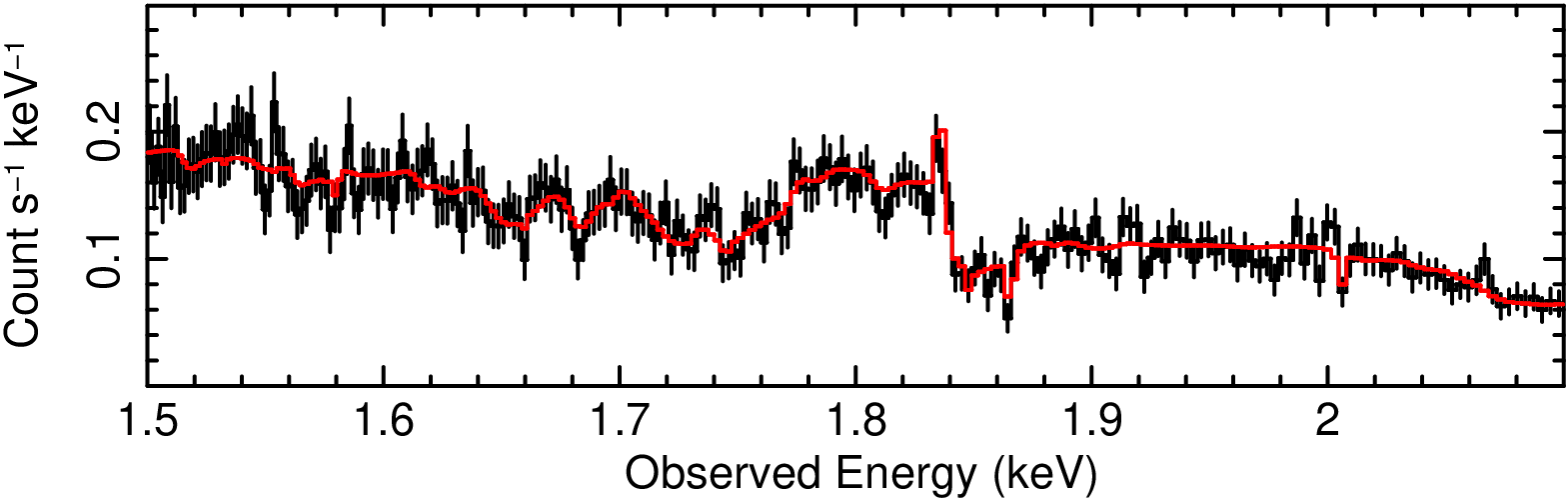}}
\end{center}
\caption{The HETG spectrum from 0.5--2.0\,keV with our best-fitting warm absorber model super-imposed. The emission lines are modelled with simple Gaussians. All spectra are binned up by a factor of 2 for clarity and are plotted in the observed frame. Note that an additional Gaussian with a negative flux was included in the fit to model the excess residuals at $\sim$666\,eV (see Section 5.1 for further details). We do note that some unmodelled residuals remain in the data at $\sim$0.87\,keV; these are likely due to inner-shell Ne transitions (e.g. Ne\,\textsc{viii}) which are not included in the \textsc{xstar} models.} 
\end{figure*}

\begin{figure}
\begin{center}
\rotatebox{-90}{\includegraphics[width=6cm]{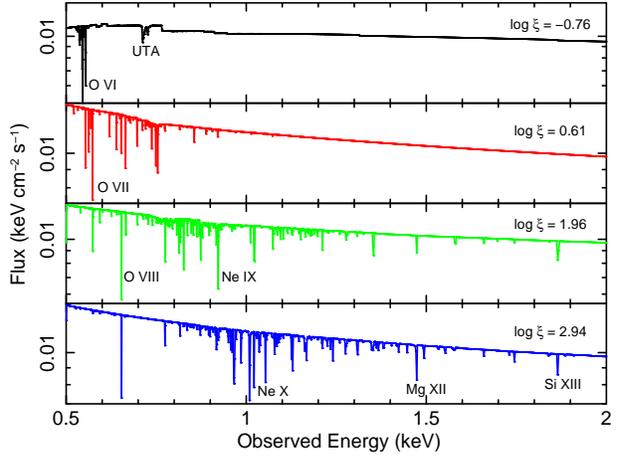}}
\end{center}
\caption{Plot showing the contribution of each of the four warm absorber zones to the power-law continuum. The lowest to highest ionisation zones (zones 1 to 4) are shown from the upper to lower panels respectively. The ions contributing to the most significant absorption lines are marked on the plot.}
\end{figure}

\begin{table}
\centering
\caption{The best-fitting parameters for the four individual zones of absorption in the HETG data. $^{a}$\textsc{xstar} zone: $N_{\rm H}$, column density; $\xi$, ionisation parameter; $v_{\rm out}$, outflow velocity. $^{b}$Partial-coverer: $f_{\rm cov}$, covering fraction. Zones 1, 3 and 4 require a turbulent velocity width of $\sigma = 200$\,km\,s$^{-1}$ whereas Zone 2 is better modelled with a turbulent velocity width of $\sigma = 500$\,km\,s$^{-1}$.}
\begin{tabular}{c c c c}
\hline
\hline
Absorption & Model & \multirow{2}{*}{Value} & \multirow{2}{*}{$\Delta C$} \\
Component & Parameter & & \\
\hline
\multirow{3}{*}{Zone 1$^{a}$} & $N_{\rm H}$\,(cm$^{-2}$) & $(3.06^{+1.04}_{-0.78}) \times 10^{20}$ & \multirow{3}{*}{149} \\
& log\,$\xi$ & $-(0.86^{+0.14}_{-0.30})$ & \\
& $v_{\rm out}$\,(km\,s$^{-1}$) & $-(180\pm100)$ & \\
\multirow{3}{*}{Zone 2$^{a}$} & $N_{\rm H}$\,(cm$^{-2}$) & $(1.52^{+0.54}_{-0.37}) \times 10^{20}$ & \multirow{3}{*}{56} \\
& log\,$\xi$ & $0.60^{+0.30}_{-0.22}$ & \\
& $v_{\rm out}$\,(km\,s$^{-1}$) & $-(220^{+40}_{-60})$ & \\
\multirow{3}{*}{Zone 3a$^{a}$} & $N_{\rm H}$\,(cm$^{-2}$) & $(1.10^{+0.84}_{-0.51}) \times 10^{21}$ & \multirow{3}{*}{106} \\
& log\,$\xi$ & $2.16^{+0.09}_{-0.21}$ & \\
& $v_{\rm out}$\,(km\,s$^{-1}$) & $-(550\pm60)$ & \\
\multirow{3}{*}{Zone 3b$^{a}$} & $N_{\rm H}$\,(cm$^{-2}$) & $(4.96^{+2.26}_{-1.48}) \times 10^{20}$ & \multirow{3}{*}{143} \\
& log\,$\xi$ & $1.96^{+0.17}_{-0.06}$ & \\
& $v_{\rm out}$\,(km\,s$^{-1}$) & $-(820\pm30)$ & \\
\multirow{3}{*}{Zone 4$^{a}$} & $N_{\rm H}$\,(cm$^{-2}$) & $(2.74^{+0.58}_{-0.66}) \times 10^{21}$ & \multirow{3}{*}{127} \\
& log\,$\xi$ & $2.97^{+0.10}_{-0.05}$ & \\
& $v_{\rm out}$\,(km\,s$^{-1}$) & $-(710^{+20}_{-40})$ & \\
\hline
\multirow{3}{*}{Partial-Coverer$^{b}$} & $N_{\rm H}$\,(cm$^{-2}$) & $(2.00^{+0.68}_{-0.53}) \times 10^{23}$ & \multirow{3}{*}{286} \\
& log\,$\xi$ & $< 2.40$ & \\
& $f_{\rm cov}$ & 30\% & \\
\hline
\end{tabular}
\end{table}

\subsubsection{Comparison with Recent RGS Data}

The warm absorber in NGC 4051 was also recently studied by Pounds \& Vaughan (2010) through {\sl XMM-Newton} RGS observations in May--June 2009. The data revealed significant absorption features primarily from H-like and He-like ions of C, N, O, Ne, Mg, Si, Ar and Fe. When modelled with \textsc{xstar}, the RGS data appear to require five separate line-of-sight outflow velocity components ranging in velocity from $v_{\rm out} \sim -500$\,km\,s$^{-1}$ to $v_{\rm out} \sim -30\,000$\,km\,s$^{-1}$. With our HETG data, we are able to confirm here the presence of the lowest velocity zone (on the order of a few hundred km\,s$^{-1}$) and also the third zone ($v_{\rm out} \sim -6\,000$\,km\,s$^{-1}$) which can clearly be seen at Fe\,K with the {\sl Suzaku} XIS (see Section 6.3 for more details). However, we find no requirement for the highest velocity zone at Fe\,K ($v_{\rm out} \sim -30\,000$\,km\,s$^{-1}$; Section 6.3); nor do we find any evidence for the second and fourth intermediate zones claimed by Pounds \& Vaughan (2010) which have velocities on the order of $v_{\rm out} \sim -4\,000$\,km\,s$^{-1}$ and $v_{\rm out} \sim - 9\,000$\,km\,s$^{-1}$ respectively. \\

Pounds \& Vaughan (2010) report that the O\,\textsc{vii} $1s$--$2p$ resonance transition appears to be detected with three separate velocity components ($-400$\,km\,s$^{-1}$; $-4\,000$\,km\,s$^{-1}$; $-6\,000$\,km\,s$^{-1}$). However, we are only able to confirm the lowest velocity component with the HETG for this transition (see Figure 7; top panel). Pounds \& Vaughan (2010) also report four separate velocity components for the O\,\textsc{viii} Ly$\alpha$ absorption line. However, we are again only able to confidently confirm the lowest velocity component in the HETG data (see Figure 7; lower panel). Interestingly, we do observe additional residuals at $\sim$666\,eV in the HETG spectrum which could potentially be the signature of a higher velocity component of the O\,\textsc{viii} Ly$\alpha$ transition with $v_{\rm out} \sim -6\,000$\,km\,s$^{-1}$ (Figure 7; lower panel), consistent with the value of $v_{\rm out} \sim -5\,600$\,km\,s$^{-1}$ found in the RGS spectrum (we note that we can confirm this velocity component at Fe\,K with the {\sl Suzaku} XIS; see Section 6.3). However, as this residual feature coincides with the expected energy of the $1s$--$3p$ transition from O\,\textsc{vii}, we may also be able to account for this with a slightly higher turbulence velocity ($\sigma = 500$\,km\,s$^{-1}$) in our \textsc{xstar} model (as mentioned in Section 5.1) so its interpretation remains open. We also note that we find no requirement at all here for the further two velocity components associated with the O\,\textsc{viii} transition at $v_{\rm out} \sim -4\,100$\,km\,s$^{-1}$ and $v_{\rm out} \sim -9\,000$\,km\,s$^{-1}$ in the RGS data. \\

\begin{figure}
\begin{center}
\rotatebox{-90}{\includegraphics[width=4.5cm]{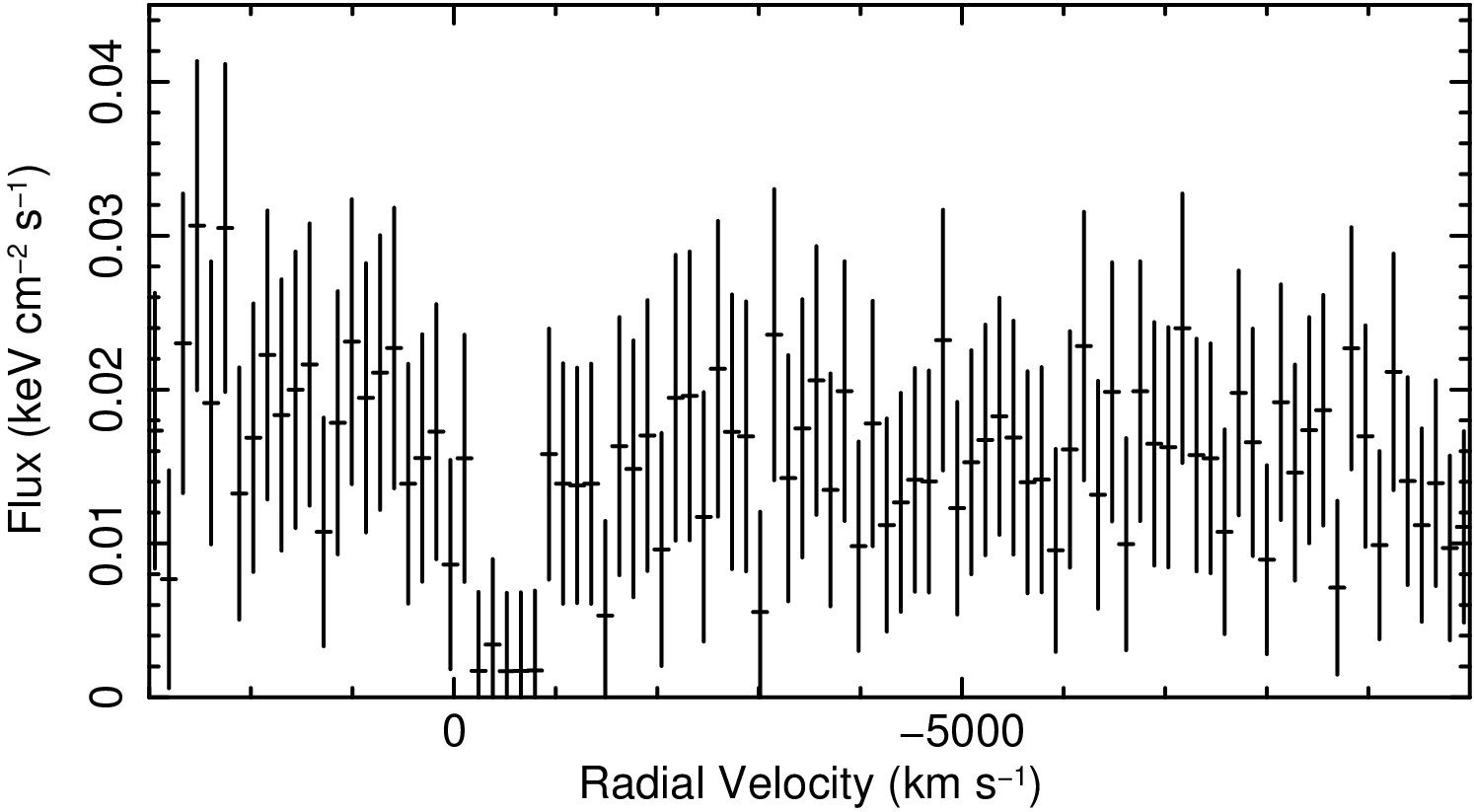}}
\rotatebox{-90}{\includegraphics[width=4.5cm]{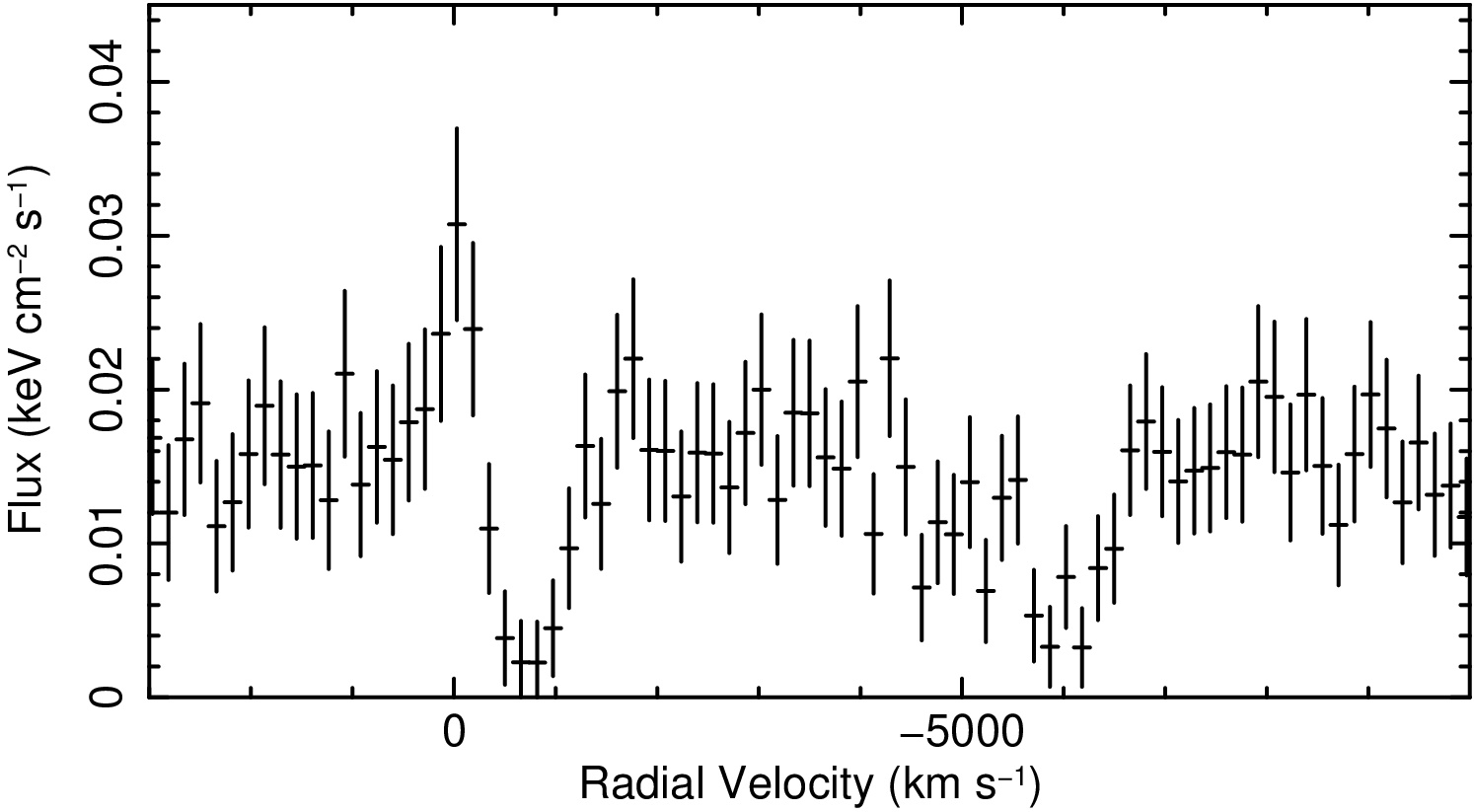}}
\end{center}
\caption{Upper panel: Plot showing a radial velocity profile centred on the O\,\textsc{vii} $1s$--$2p$ resonance transition at $\sim$574\,eV. Note that we see evidence for an associated absorption component with an outflow velocity of a few hundred km\,s$^{-1}$ but we find no requirement for the higher velocity components reported by Pounds \& Vaughan (2010). Lower panel: Plot showing a radial velocity profile centred on the O\,\textsc{viii} Ly$\alpha$ transition at $\sim$653\,eV. Again, we see evidence for the lowest velocity absorption component with $v_{\rm out} < 1,000$\,km\,s$^{-1}$. We do see an absorption feature which may correspond to O\,\textsc{viii} Ly$\alpha$ blueshifted by $v_{\rm out} \sim -6\,000$\,km\,s$^{-1}$; however, we may also be able to associate this feature with the $1s$--$3p$ transition from O\,\textsc{vii}.}
\end{figure}

Furthermore, Pounds \& Vaughan (2010) also report that the Ne\,\textsc{x} and Mg\,\textsc{xii} Ly$\alpha$ absorption lines are blueshifted on the order of $v_{\rm out} \sim -7\,000$\,km\,s$^{-1}$ in the RGS data whereas we only find evidence for these components with $v_{\rm out} < 1\,000$\,km\,s$^{-1}$ with the {\sl Chandra} HETG. Therefore, we are only able to confirm two of the five velocity components associated with the {\sl XMM-Newton} data, although we cannot rule out the possibility that the absorber may have significantly varied between observations.

\subsection{The Fe\,K Complex}

A ratio plot of the residuals to the absorbed baseline HETG continuum from 5.5--7.5\,keV is shown in Figure 8. Line emission is clearly present suggestive of a significant Fe\,K complex. The most prominent line appears to be the Fe\,K$\alpha$ fluorescence line from near-neutral material at $E_{\rm line} = 6.41\pm0.03$\,keV. Modelling this with a Gaussian reveals that this line has an intrinsic width of $\sigma = 0.12^{+0.04}_{-0.01}$\,keV (FWHM $= 13\,000^{+4\,000}_{-1\,000}$\,km\,s$^{-1}$) and an associated equivalent width of $EW = 185^{+58}_{-19}$\,eV. The addition of this line significantly improved the fit by $\Delta C = 99$. However, the fit was improved further by incorporating an additional unresolved narrow component ($\sigma = 0$\,eV) centered at 6.40\,keV to model any Fe\,K$\alpha$ emission from distant material. This component has an equivalent width of $EW = 50^{+18}_{-16}$\,eV and resulted in an improvement to the fit statistic of $\Delta C = 19$. Upon the addition of this component, the parameters of the underlying broader K$\alpha$ component adjusted slightly to compensate but were still consistent within the errors (see Table 4 for the final best-fitting values). We did also test for the presence of a Compton shoulder by including an additional Gaussian centred around $\sim$6.3\,keV. This feature is statistically unrequired by the data and an upper limit on the equivalent width is found to be $EW < 6$\,eV. Including this component did not result in any changes to the parameters of the broad and narrow Fe\,K$\alpha$ components at $\sim$6.4\,keV. So although the near-neutral Fe\,K$\alpha$ emission appeared to then be well modelled, further residuals were still present indicating additional K-shell emission from ionised Fe. \\

\begin{figure}
\begin{center}
\rotatebox{-90}{\includegraphics[width=6cm]{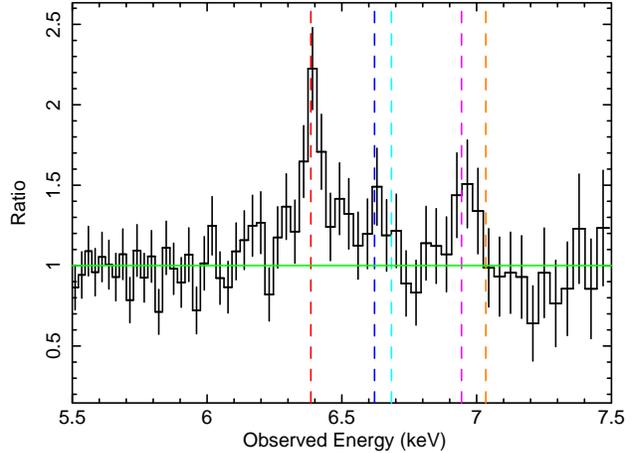}}
\end{center}
\caption{Ratio plot showing the residuals of the HEG data from 5.5--7.5\,keV to a power-law absorbed by Galactic hydrogen and a partial-coverer (as described in Section 5.1). Significant Fe\,K line emission is clearly present. The data are binned by a factor of 2 for clarity. The vertical dotted lines show the expected line energies of, from left to right, Fe\,\textsc{i-xvii}\,K$\alpha$ (red), Fe\,\textsc{xxv} forbidden (blue), Fe\,\textsc{xxv}\,K$\alpha$ resonance (cyan), Fe\,\textsc{xxvi} Ly$\alpha$ (magenta) and Fe\,\textsc{i-xvii}\,K$\beta$ (orange) in the observed frame.} 
\end{figure}

We adopted additional Gaussians to model the remaining emission. A line at $E_{\rm line} = 6.97\pm0.02$\,keV is required by the data with an intrinsic width of $\sigma = 32^{+34}_{-17}$\,eV (FWHM $=3\,100^{+3\,400}_{-1\,600}$\,km\,s$^{-1}$) and an associated equivalent width of $EW = 63^{+32}_{-29}$\,eV. This line is most likely due to the $1s$--$2p$ doublet from hydrogen-like iron (Fe\,\textsc{xxvi}) and corresponds to an improvement in the fit statistic of $\Delta C = 15$. A third component with a centroid energy of $E_{\rm line} = 6.64\pm0.02$\,keV is also required by the data. This line has an intrinsic width of $\sigma < 52$\,eV (FWHM $< 5\,400$\,km\,s$^{-1}$) and corresponds to an equivalent width of $EW = 18^{+16}_{-13}$\,eV. This line improves the fit statistic by $\Delta C = 6$ (i.e. only $\sim$95 per cent significant) and is most likely associated with being the forbidden transition of the helium-like iron (Fe\,\textsc{xxv}) triplet at 6.636\,keV, suggesting that this emission line may be consistent with an origin in photo-ionised gas (Bautista \& Kallman 2000; Porquet \& Dubau 2000). No further emission or absorption features were observed around the Fe\,K complex in the HETG data and the best-fitting values of the emission lines are shown in Table 4. The overall fit statistic corresponds to $C / d.o.f. = 4779 / 4426$. \\

\begin{table*}
\centering
\caption{The best-fitting Fe\,K parameters from the {\sl Chandra} HETG. All values are given in the rest frame of the source. See Section 5.2 for details.}
\begin{tabular}{l c c c c c c}
\hline\hline
Line ID & $E_{\rm line}$ & $\sigma$ & FWHM & $F_{\rm line}$ & EW & $\Delta C$ \\ [0.5ex]
& (keV) & (eV) & (km\,s$^{-1}$) & ($\times 10^{-5}$\,photon\,cm$^{-2}$\,s$^{-1}$) & (eV) & \\
\hline
Fe\,K$\alpha_{\rm broad}$ & $6.43\pm0.06$ & $150^{+60}_{-43}$ & $16\,000^{+7\,000}_{-4\,000}$ & $3.17^{+0.91}_{-0.89}$ & $156^{+45}_{-44}$ & 36 \\
Fe\,K$\alpha_{\rm narrow}$ & $6.40\pm0.01$ & 0$_{\rm fixed}$ & $\sim$1\,000 & $1.16^{+0.41}_{-0.37}$ & $50^{+18}_{-16}$ & 25  \\
$[$Fe\,\textsc{xxv}$]$ $1s$--$2s$ & $6.64\pm0.02$ & $< 52$ & $< 5\,400$ & $0.43^{+0.37}_{-0.30}$ & $19^{+16}_{-13}$ & 6 \\
Fe\,\textsc{xxvi} $1s$--$2p$ & $6.97\pm0.02$ & $32^{+34}_{-17}$ & $3\,100^{+3\,400}_{-1\,600}$ & $1.01^{+0.51}_{-0.47}$ & $63^{+32}_{-29}$ & 15 \\
\hline
\end{tabular}
\end{table*}

With the parameterisation of the baseline continuum, the warm absorber and the Fe\,K components complete, we consider this to be our final best-fitting model to the time-averaged {\sl Chandra} HETG data. We then proceed to apply this model to the 2008 contemporaneous {\sl Suzaku} data.

\section{Suzaku Spectral Analysis}

We begin by considering the time-averaged 2008 {\sl Suzaku} spectrum. In all subsequent fits, we use data taken in the 0.6--10.0 and 15.0--50.0\,keV energy bands for the XIS and HXD respectively. The cross-normalisation between the HXD PIN and the XIS detectors was accounted for by the addition of a fixed constant component at a value of 1.18 for the HXD nominal pointing position (2005 observation) and 1.16 for the XIS nominal pointing position (2008 observations); values derived using {\sl Suzaku} observations of the Crab (Ishida, Suzuki \& Someya 2007\footnote{ftp://legacy.gsfc.nasa.gov/suzaku/doc/xrt/suzakumemo-2007-11.pdf}). We also ignored all data from 1.7--2.0\,keV so as to avoid any contamination from the absorption edge due to silicon in the detectors. Since the XIS data were binned up at the HWHM of the resolution of the detector, we included a 2 per cent systematic error in all fits to account for the high statistical weight of the bins at low energies.

\subsection{The Broad-Band Suzaku Model}

We firstly applied the best-fitting HETG continuum model (from Section 5) consisting of a power-law continuum, a partial-coverer, the fully-covering warm absorber (zones 1--4), Fe\,K emission lines and a black body to parameterise the soft excess to the time-averaged 2008 broad-band {\sl Suzaku} data from 0.6--50.0\,keV when the source was observed to be in a period of high flux ($F_{\rm 0.5-10.0} = 5.03 \times 10^{-11}$\,erg\,cm$^{-2}$\,s$^{-1}$). A historical lightcurve of NGC 4051 in the X-ray band is shown in Figure 9 where it can be seen that the 2008 observations were made during a period of relatively high flux for this source. Since the 2008 data were obtained in two separate observations due to scheduling constraints, we initially only considered the much longer $\sim$275\,ks exposure. We included all of the soft X-ray, fully covering absorption zones required by the HETG data (including the soft emission lines) and fixed the parameters at the best-fitting values listed in Table 3 since the 2008 {\sl Chandra} and {\sl Suzaku} data were quasi-simultaneous. This resulted in a very poor fit with $\chi^{2} / d.o.f. = 1351 / 188$. Allowing the model to renormalise slightly improved the fit but still returned a poor fit statistic of $\chi^{2} / d.o.f. = 1165 / 188$ with the main contribution to the residuals arising from the significant hard excess seen at energies \textgreater 10\,keV. This hard excess can be seen in Figure 13 (see Section 6.5) where broad-band {\sl Suzaku} spectrum is shown (in red) unfolded against a power-law with $\Gamma = 2$.

\begin{figure}
\begin{center}
\rotatebox{-90}{\includegraphics[width=6cm]{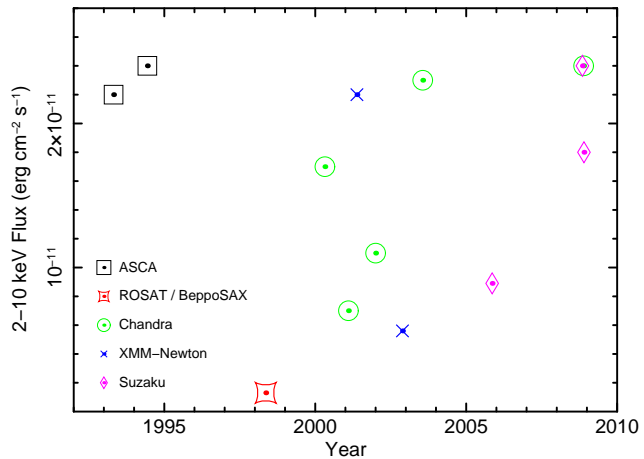}}
\end{center}
\caption{A historical lightcurve showing the 2--10\,keV flux from all observations of NGC 4051 with {\sl ASCA}, {\sl BeppoSAX} (simultaneously with {\sl ROSAT}), {\sl Chandra}, {\sl XMM-Newton} and {\sl Suzaku}. The lightcurve shows that the flux of this source varies significantly over time and that the 2008 {\sl Chandra} and {\sl Suzaku} observations that we describe here (top-right) were made during a period of high flux. Note that the 2005 {\sl Suzaku} data modelled in Section 6.5 as well as previously by Terashima et al. (2009) were taken during a period of much lower flux. Data for the lightcurve were obtained from Guainazzi et al. (1996), Guainazzi et al. (1998), Uttley et al. (1999), Collinge et al. (2001), Uttley et al. (2003), Ogle et al. (2004), Steenbrugge et al. (2009), Terashima et al. (2009), Shu, Yaqoob \& Wang (2010) and Turner et al. (2010).}
\end{figure}

\subsection{The Suzaku Fe\,K Profile}

In order to accurately analyse the {\sl Suzaku} Fe\,K profile, it was therefore important to firstly parameterise the broad-band continuum emission using both the XIS and HXD data. To do this, we modelled the hard excess by including the \textsc{pexrav} model (Magdziarz \& Zdziarski 1995); an additive component incorporating the reflected continuum from a neutral slab. We tied the photon index and the normalisation of the unabsorbed power-law continuum to that of the power-law component incident upon the reflector and fixed the elemental abundances to Solar (Anders \& Grevesse 1989). We fixed the cosine of the inclination angle of the source to 0.87 (corresponding to 30$^{\circ}$) and tied the folding energy to the cutoff energy of the power-law at 300\,keV, consistent with no cut-off in the HXD PIN band. We found that the combination of the \textsc{pexrav} and blackbody models (to account for the hard and soft excess respectively) with the partial-coverer from the HETG model (to account for the additional spectral curvature) was able to smoothly paramaterise the broad-band continuum well. \\

We then began to parameterise the Fe\,K profile by modelling the most prominent emission line corresponding to K$\alpha$ emission from near-neutral iron with a Gaussian. This emission line is found to have a centroid energy of $E_{\rm line} = 6.41\pm0.01$\,keV in the {\sl Suzaku} data and an intrinsic width of $\sigma < 53$\,eV (FWHM \textless 5\,700\,km\,s$^{-1}$). It also has an associated equivalent width of $EW = 75^{+14}_{-9}$\,eV against the observed continuum. The inclusion of this component improves the fit statistic by $\Delta \chi^{2} = 228$. The fit was further improved ($\Delta \chi^{2} = 12$) by including a second line to model the forbidden transition from Fe\,\textsc{xxv} found in Section 5.2 which appears to be manifested in the slight blue wing of the Fe\,K$\alpha$ line shown in Figure 10. The line width was fixed at $\sigma = 0$\,eV (intrinsically narrow) and although the centroid energy was difficult to constrain, it was found to have a best-fitting value of $E_{\rm line} = 6.62^{+0.09}_{-0.02}$\,keV. The best-fitting values of the line flux and associated equivalent width ($F_{\rm line} = 5.51^{+1.29}_{-2.39} \times 10^{-6}$\,photons\,cm$^{-2}$\,s$^{-1}$ and $EW = 25^{+6}_{-11}$\,eV respectively) are also largely consistent with the corresponding values found in the {\sl Chandra} HETG data. \\

\begin{figure}
\begin{center}
\rotatebox{-90}{\includegraphics[width=6cm]{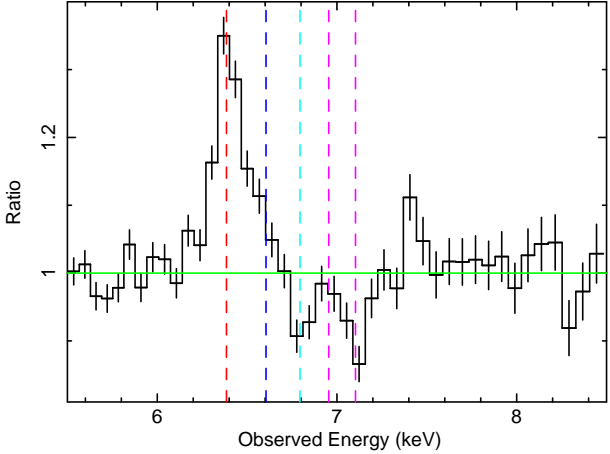}}
\end{center}
\caption{Plot showing the ratio of the data-to-model residuals of the 2008 {\sl Suzaku} XIS data in the Fe\,K band in the observed frame. The dotted lines show the best-fitting centroid energies of the five significant emission / absorption lines required by the data: from left to right, neutral Fe\,K$\alpha$ emission at 6.41\,keV (red), Fe\,\textsc{xxv} forbidden emission at 6.62\,keV (blue), blueshifted absorption most likely from Fe\,\textsc{xxv} at 6.81\,keV (cyan) and emission and blueshifted absorption from Fe\,\textsc{xxvi} Ly$\alpha$ at 6.97\,keV and 7.12\,keV respectively (both magenta).} 
\end{figure}

We also included a third line to model the emission from the Fe\,\textsc{xxvi} $1s$--$2p$ Ly$\alpha$ transition. Including an additional line fixed at the best-fitting value found in the HETG data of $E_{\rm line} = 6.97$\,keV with an intrinsically narrow width of $\sigma = 0$\,eV (see Section 5.2 and Table 4) is marginally required by the data ($\Delta \chi^{2} \sim 5$) and corresponds to an equivalent width of $EW = 9.9^{+9.3}_{-7.8}$\,eV against the observed continuum. This line appears to be less significant in the {\sl Suzaku} XIS data than in the {\sl Chandra} HETG data possibly due to the strength of two significant absorption lines with blueshifted centroid energies of $\sim$6.8\,keV and $\sim$7.1\,keV (in the observed frame) possibly masking some of the emission from the $1s$--$2p$ transition of H-like Fe in the XIS data (see Section 6.3). We also note that a slight positive residual can be observed at $\sim$7.5\,keV (see Figure 10) which is likely due to K$\alpha$ fluorescence from near-neutral Ni which has rest-frame centroid energy of $\sim$7.47\,keV. However, modelling this feature with a Gaussian is statistically unrequired by the data. \\

Finally, although no neutral Fe\,K$\beta$ emission was statistically required by the data ($\Delta \chi^{2} \sim 2$), we modelled it for consistency with the addition of a further Gaussian. We fixed the line energy at 7.06\,keV, tied the intrinsic width to that of the corresponding Fe\,K$\alpha$ line and fixed the line flux at 13 per cent of the K$\alpha$ flux, consistent with the theoretical flux ratio for near-neutral iron (Kaastra \& Mewe 1993). This line is unrequired by the data but for consistency we keep the line modelled in all subsequent fits.

\subsection{Highly Ionised Absorption}

Upon modelling the Fe\,K emission it can be observed that significant negative residuals are apparent in the data at energies $\sim$7\,keV suggestive of the presence of highly ionised absorption lines. At a first glance, the inferred blueshift of these lines could suggest that these lines are the absorption signature of a high velocity outflow, such as that detected by Pounds et al. (2004a). A plot of the ratio of the data-to-model residuals in the 2008 {\sl Suzaku} XIS data is shown in Figure 10. We initially parameterised these absorption lines with Gaussians with negative fluxes. The first line is found to have a centroid energy of $E_{\rm line} = 6.81^{+0.04}_{-0.05}$\,keV, improving the fit by $\Delta \chi^{2} = 26$ (also see Figure 11). Its intrinsic width was fixed at a value of $\sigma = 0$\,eV. This absorption line has an associated equivalent width of $EW = -(28^{+31}_{-11})$\,eV and is most likely due to absorption from the Fe\,\textsc{xxv} resonance line at 6.70\,keV. If so, the blueshift of the line would suggest that the absorbing material is outflowing with a velocity of $v_{\rm out} = -(5\,500)^{+1\,800}_{-2\,100}$\,km\,s$^{-1}$ relative to the host galaxy, consistent with the value found by Terashima et al. (2009) and Pounds \& Vaughan (2010). The second line appears to correspond to a centroid energy of $E_{\rm line} = 7.12\pm0.04$\,keV with a corresponding equivalent width of $EW = -(43^{+9}_{-11})$\,eV. The addition of this line greatly improves the fit statistic by $\Delta \chi^{2} = 56$ (also see Figure 11). Again the intrinsic width of the line was fixed at $\sigma = 0$\,eV. If this line is associated with the $1s$--$2p$ Ly$\alpha$ doublet from Fe\,\textsc{xxvi} at $\sim$6.97\,keV then it would correspond to an outflow velocity of $v_{\rm out} = -(7\,500^{+1\,600}_{-1\,700})$\,km\,s$^{-1}$ again consistent with the findings of Pounds et al. (2004a) and Terashima et al. (2009). \\

\begin{figure}
\begin{center}
\rotatebox{-90}{\includegraphics[width=6cm]{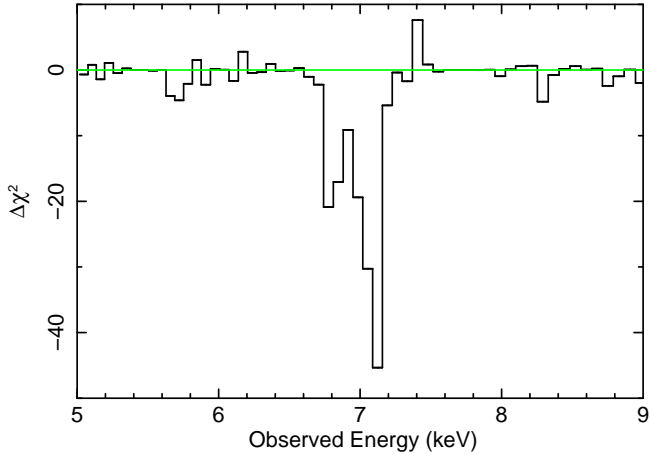}}
\end{center}
\caption{Plot (in the observed frame) showing the significant contribution to the $\chi^{2}$ value of the two highly ionised absorption lines at $\sim$6.8 and $\sim$7.1\,keV most likely originating from Fe\,\textsc{xxv} and Fe\,\textsc{xxvi} respectively. The Fe\,K emission lines are already modelled, as detailed in Section 6.2.} 
\end{figure}

After parameterising the absorption lines with simple Gaussians, we attempted to self-consistently model the lines with the addition of a photo-ionised grid of absorbing gas using the \textsc{xstar} code of Kallman \& Bautista (2001). We adopted a turbulence velocity of $\sigma = 3\,000$\,km\,s$^{-1}$ so as not to saturate the absorption lines and to explain their high observed equivalent widths. We find that the addition of this zone of gas improves the fit by $\Delta \chi^{2} = 113$ and requires a best-fitting column density of $N_{\rm H} = 8.4^{+1.9}_{-2.0} \times 10^{22}$\,cm$^{-2}$ and a best-fitting value for the ionisation parameter of log\,$\xi = 4.1^{+0.2}_{-0.1}$. The blueshift of the zone corresponds to an outflow velocity of $v_{\rm out} = -(5\,800^{+860}_{-1\,200})$\,km\,s$^{-1}$ ($\sim$$-0.02c$), consistent with the values found by Pounds et al. (2004a) and Terashima et al. (2009) for the highly ionised outflow. We note that no further zones of absorption are statistically required by the data to model the absorption at Fe\,K and that we do not see any evidence of the higher velocity ($v_{\rm out} \sim -30\,000$\,km\,s$^{-1}$) component reported by Pounds \& Vaughan (2010) (see Figure 12). Indeed, by fixing two additional Gaussians centred on energies of 7.5 and 7.7\,keV (corresponding to the observed energies reported by Pounds \& Vaughan 2010), we note that these are unrequired by the data and have upper limits on their equivalent widths of $EW < 7$\,eV and $EW < 13$\,eV respectively. \\

\begin{figure}
\begin{center}
\rotatebox{-90}{\includegraphics[width=5cm]{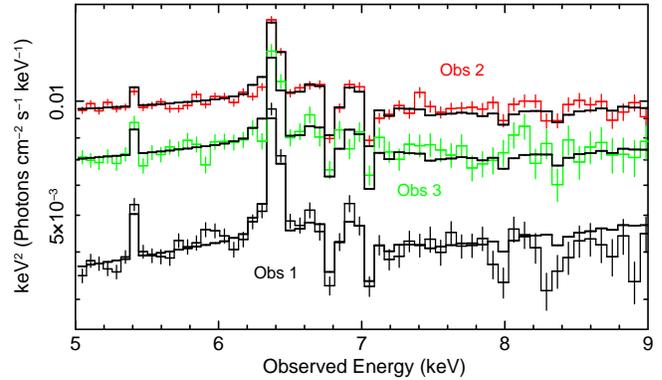}}
\end{center}
\caption{Plot showing the Fe\,K region for the three {\sl Suzaku} observations of NGC 4051. We find strong evidence for a highly ionised outflow with $v_{\rm out} \sim -6\,000$\,km\,s$^{-1}$ due to the two absorption lines at $\sim$6.8 and $\sim$7.1\,keV which we associate with the $1s$--$2p$ transitions from Fe\,\textsc{xxv} and Fe\,\textsc{xxvi} respectively. However, we see no evidence for the much faster component with $v_{\rm out} \sim -30\,000$\,km\,s$^{-1}$ reported by Pounds \& Vaughan (2010) which they associate with the presence of two additional absorption lines at $\sim$7.5--7.7\,keV. We note that slight residuals at $\sim$8.0 and $\sim$8.3\,keV can be seen in the 2005 data; these are likely due to the $1s$--$3p$ transitions of Fe\,\textsc{xxv} and Fe\,\textsc{xxvi} respectively but are not statistically significant. The 2005 data are shown in black and the two separate 2008 observations are shown in red and green corresponding to the 275 and 78\,ks exposures respectively.} 
\end{figure}

Upon modelling the Fe\,K emission and the highly ionised absorption, the reflection scaling factor of the \textsc{pexrav} component was found to have a best-fitting value of $R = 1.23^{+0.13}_{-0.06}$. 
We note that a sub-Solar Fe abundance appears to be preferred by the data with the value dropping to $A_{\rm Fe} = 0.37^{+0.08}_{-0.11}$ times Solar. However, this could simply be due to the limitations of fitting the spectrum with a simple neutral slab. We also note that the ratio of the normalisations of the absorbed to total power-law components at 1\,keV ($\sim$0.3) appears to be consistent with the best-fitting values obtained from the HETG spectrum (i.e. still suggestive of a $\sim$30 per cent covering fraction for the partial-covering zone; also see Section 5.1). The final best-fitting parameters of the broad-band model to the {\sl Suzaku} XIS/HXD data are shown in Table 5. The overall fit statistic corresponds to $\chi^{2} / d.o.f. = 218 / 187$. \\

Finally, as a consistency check, we returned to the HETG data to test for the presence of the two highly ionised absorption lines which we detect with {\sl Suzaku}. Taking the best-fitting HETG model from Section 5, we added two Gaussian profiles with negative fluxes fixing the centroid energies at $E_{\rm line} = 6.81$\,keV and $E_{\rm line} = 7.12$\,keV respectively and the intrinsic width of the two lines at $\sigma = 0$\,eV, consistent with the {\sl Suzaku} data. Neither of the two lines were statistically required by the HETG data but we find upper limits on the magnitudes of the equivalent widths of $EW < 21$\,eV and $EW < 41$\,eV for the two respective lines, indicating that they are consistent with the values that we find with {\sl Suzaku}.

\begin{table}
\centering
\caption{Table showing the best-fitting rest-frame parameters of the broad-band {\sl Suzaku} XIS+HXD model described in Section 6. $^{a}$Primary power-law continuum: $\Gamma$, photon index; Normalisation. $^{b}$Absorbed power-law. $^{c}$\textsc{pexrav} neutral reflector: $R$, reflection scaling factor; $A_{\rm Fe}$, iron abundance with respect to Solar. $^{d}$Blackbody: $kT$, temperature; Normalisation defined as $L_{\rm 39} / D^{2}_{\rm 10}$ where $L_{\rm 39}$ is the source luminosity in units of $10^{39}$\,erg\,s$^{-1}$ and $D_{\rm 10}$ is the distance to the source in units of 10\,kpc. $^{e}$Emission/Absorption lines: $E_{\rm line}$, centroid energy; $\sigma$, intrinsic width; $F_{\rm line}$, line flux; $EW$, equivalent width. $^{f}$Highly ionised absorption: $N_{\rm H}$, column density; $\xi$, ionisation parameter; $v_{\rm out}$, velocity shift. The symbol $l$ signifies that the parameter is linked to all parameters with the corresponding number. Note that the high $\xi$ \textsc{xstar} zone replaced the highly ionised absorption lines and they were not modelled simultaneously.}
\begin{tabular}{l c c c}
\hline\hline
Component & Parameter & Value & $\Delta \chi^{2}$ \\
\hline
\multirow{2}{*}{Power-law$^{a}$} & $\Gamma$ & $2.20\pm0.01^{^{l_{\rm 1}}}$ & \\
& Norm.\,(ph\,cm$^{-2}$\,s$^{-1}$) & $(1.14\pm0.01) \times 10^{-2}$ & \\
\hline
\multirow{2}{*}{Power-law$^{b}_{\rm abs.}$} & $\Gamma$ & $l_{\rm 1}$ & \\
& Norm.\,(ph\,cm$^{-2}$\,s$^{-1}$) & $(5.19^{+0.56}_{-0.69}) \times 10^{-3}$ & \\
\hline
\multirow{2}{*}{\textsc{pexrav}$^{c}$} & $R$ & $1.23^{+0.13}_{-0.06}$ & \\
& $A_{\rm Fe}$ & $0.37^{+0.08}_{-0.11}$ & \\
\hline
\multirow{2}{*}{Blackbody$^{d}$} & $kT$\,(keV) & $0.11\pm0.01$ & \\
& Norm.\,($L_{\rm 39} / D^{2}_{\rm 10}$) & $(2.34^{+0.18}_{-0.26}) \times 10^{-4}$ & \\
\hline
\multirow{4}{*}{Fe\,K$\alpha^{e}$} & $E_{\rm line}$\,(keV) & $6.41\pm0.01$ & \multirow{4}{*}{228} \\
& $\sigma$\,(eV) & $< 53$ & \\
& $F_{\rm line}$\,(ph\,cm$^{-2}$\,s$^{-1}$) & $(1.69^{+0.32}_{-0.20}) \times 10^{-5}$ & \\
& EW\,(eV) & $75^{+14}_{-9}$ & \\
& & & \\
\multirow{4}{*}{Fe\,\textsc{xxv}$^{e}$} & $E_{\rm line}$\,(keV) & $6.62^{+0.09}_{-0.02}$ & \multirow{4}{*}{12} \\
& $\sigma$\,(eV) & $0_{\rm fixed}$ & \\
& $F_{\rm line}$\,(ph\,cm$^{-2}$\,s$^{-1}$) & $(5.51^{+1.29}_{-2.39}) \times 10^{-6}$ & \\
& EW\,(eV) & $25^{+6}_{-11}$ & \\
& & & \\
\multirow{4}{*}{Fe\,\textsc{xxvi}$^{e}$} & $E_{\rm line}$\,(keV) & $6.97_{\rm fixed}$ & \multirow{4}{*}{5} \\
& $\sigma$\,(eV) & $0_{\rm fixed}$ & \\
& $F_{\rm line}$\,(ph\,cm$^{-2}$\,s$^{-1}$) & $(1.93^{+1.82}_{-1.52}) \times 10^{-6}$ & \\
& EW\,(eV) & $10^{+9}_{-8}$ & \\
\hline
\multirow{4}{*}{Fe\,\textsc{xxv}$^{e}_{\rm abs.}$} & $E_{\rm line}$\,(keV) & $6.81^{+0.04}_{-0.05}$ & \multirow{4}{*}{26} \\
& $\sigma$\,(eV) & $0_{\rm fixed}$ & \\
& $F_{\rm line}$\,(ph\,cm$^{-2}$\,s$^{-1}$) & $-(5.14^{+5.74}_{-1.95}) \times 10^{-6}$ & \\
& EW\,(eV) & $-(28^{+31}_{-11})$ & \\
& & & \\
\multirow{4}{*}{Fe\,\textsc{xxvi}$^{e}_{\rm abs.}$} & $E_{\rm line}$\,(keV) & $7.12\pm0.04$ & \multirow{4}{*}{56} \\
& $\sigma$\,(eV) & $0_{\rm fixed}$ & \\
& $F_{\rm line}$\,(ph\,cm$^{-2}$\,s$^{-1}$) & $-(6.95^{+1.41}_{-1.81}) \times 10^{-6}$ & \\
& EW\,(eV) & $-(43^{+9}_{-11})$ & \\
& & & \\
\multirow{3}{*}{High $\xi$ abs.$^{f}$} & $N_{\rm H}$\,(cm$^{-2}$) & $(8.36^{+1.89}_{-2.01}) \times 10^{22}$ & \multirow{3}{*}{113} \\
& log\,$\xi$ & $4.10^{+0.18}_{-0.11}$ & \\
& $v_{\rm out}$\,(km\,s$^{-1}$) & $-(5\,800^{+860}_{-1\,200})$ & \\
\hline
\end{tabular}
\end{table}

\subsection{Absorbed Reflection Model}

Since the \textsc{pexrav} and blackbody components served as simple parameterisations of the Compton-scattered reflection hump off neutral material at energies \textgreater 10\,keV and the soft excess at energies \textless 2\,keV respectively, we attempted to model these features using a more physical approach. We began by removing the \textsc{pexrav} component, the blackbody and the three Gaussians at 6.41, 6.62 and 6.97\,keV (modelling emission from near-neutral Fe\,K$\alpha$, Fe\,\textsc{xxv} and Fe\,\textsc{xxvi} respectively). We then replaced these components with two absorbed reflection models (consistent with the full broad-band model described in Turner et al. (2010)) using the \textsc{reflionx} code of Ross \& Fabian (2005); one low $\xi$ (near-neutral) component to model the near-neutral Fe\,K$\alpha$ emission line and the hard excess and one highly ionised component to account for the spectral curvature in the soft band (i.e. the soft excess) plus the highly ionised Fe emission lines. The \textsc{reflionx} code models the emergent spectrum from a photo-ionised optically-thick slab of gas when irradiated by a power-law spectrum and consists of both the reflected continuum and line emission for the astrophysically abundant elements. It assumes a high-energy exponential cut-off of $E_{\rm cut} = 300$\,keV and uses the abundances of Anders \& Ebihara (1982). All absorption was again modelled using the \textsc{xstar} 2.1ln11 code of Kallman \& Bautista (2001) with the soft X-ray absorption fixed at the best-fitting values from the HETG data (Table 3) and the zone of highly ionised absorption detected with {\sl Suzaku} consistent with the values obtained in Section 6.3. Any soft X-ray emission lines unable to be accounted for by the ionised reflector were modelled with Gaussians as required. This model can be expressed as:

\begin{eqnarray}
F(E) & = & \textsc{wabs} \times (\textsc{xstar}_{\rm HETG} \times \textsc{xstar}_{\rm Fe\,K} \times [{\rm PL_{\rm int.}} \nonumber \\
& & + ({\rm PL} \times \textsc{xstar}_{\rm pc}) + (\textsc{reflionx}_{\rm low\,\xi} \times \textsc{xstar}) \nonumber \\
& & + (\textsc{reflionx}_{\rm high\,\xi} \times \textsc{xstar})] + {\rm GA^{\rm soft}_{\rm ems.}}),
\end{eqnarray}

where \textsc{wabs} is the absorption due to Galactic hydrogen in our line-of-sight, \textsc{xstar}$_{\rm HETG}$ corresponds to the four fully-covering soft X-ray absorption zones fixed at the best-fitting values from the HETG data (see Table 3), \textsc{xstar}$_{\rm Fe\,K}$ is the zone of highly ionised absorption (see Section 6.3), PL$_{\rm int.}$ is the intrinsic (``unabsorbed'') power-law continuum, PL $\times$ \textsc{xstar}$_{\rm pc}$ corresponds to the partial-coverer, the \textsc{reflionx} components are the near-neutral and highly ionised absorbed reflectors (each of which is absorbed by a layer of gas within \textsc{xstar}, without which the soft excess cannot be modelled; see below) and GA$^{\rm soft}_{\rm ems.}$ corresponds to the soft emission lines which the reflector is unable to account for (also see Section 4.3); primarily the forbidden transitions from O\,\textsc{vii}, Ne\,\textsc{ix} and Si\,\textsc{xiii}. These lines were fixed at the best-fitting values from the HETG data (listed in Table A1). This forms a highly complex model since were are attempting to account for a complex ionised scattering spectrum with components that cannot neccessarily account for all of the physical effects that we expect to be present. Therefore, its apparent complexity may simply be an artefact of the limitations of the current physical models available, e.g. compared to physically realistic disc wind models (see Sim et al. 2008, 2010). \\

We fixed the redshift values of the two reflectors to that of the host galaxy (i.e. $z = 0.002336$), tied the iron abundances together and tied the photon indexes to that of the incident power-law continuum. No additional velocity broadening was applied to the reflected spectrum. We find that the photon index steepens to $\Gamma = 2.50\pm0.02$. The ionisation parameter of the low $\xi$ reflector pegs at a value of $\xi = 10$\,erg\,cm\,s$^{-1}$ (the lowest value allowed by the model) whereas the highly ionised reflector modelling the soft excess requires a best-fitting value of log\,$\xi = 3.8^{+0.2}_{-0.1}$. We note that the highly ionised reflector is able to account for the soft excess due to its enhanced reflectivity in the soft band at high levels of ionisation although we stress that the soft excess cannot be modelled by the \textsc{reflionx} component alone without introducing a zone of absorption (in addition to the four warm absorbers, the partial-coverer and the Galactic absorption). The absorber in front of the reflector is responsible for producing deep bound-free edges which reduce the observed reflected flux at higher energies, thus revealing positive spectral curvature in the soft band. Such a component may be representative of scattering off a disc wind. The highly ionised reflector also removes the residuals at Fe\,K and well models the ionised emission lines from Fe\,\textsc{xxv} and Fe\,\textsc{xxvi}. This model appeared to leave very few residuals in the data and gave a good fit with $\chi^{2} / d.o.f. = 181 / 187$, a significant improvement over the phenomenological model described in Section 6.3 ($\chi^{2} / d.o.f. = 218 / 187$). The best-fitting values of the reflection components and their associated zones of absorption are shown in Table 6.

\begin{table}
\centering
\caption{Table showing the best-fitting rest-frame parameters of the broad-band {\sl Suzaku} XIS+HXD absorbed reflection model described in Section 6.4. $^{a}$Primary power-law continuum: $\Gamma$, photon index; Normalisation. $^{b}$Absorbed power-law: $N_{\rm H}$, column density; $\xi$, ionisation parameter. $^{c}$Neutral reflector: $A_{\rm Fe}$, iron abundance with respect to Solar. $^{d}$Zone of absorption. The redshift of the zone was fixed at the same value as that of the host galaxy (i.e. $z = 0.002336$). $^{e}$Ionised reflector. The symbol $p$ signifies that the parameter has pegged at the maximum / minimum value allowed by the model and the symbol $l$ signifies that the parameter is linked to all parameters with the corresponding number.}
\begin{tabular}{l c c}
\hline\hline
Component & Parameter & Value \\
\hline
\multirow{2}{*}{Power-law$^{a}$} & $\Gamma$ & $2.50\pm0.02^{^{l_{\rm 1}}}$ \\
& Norm.\,(ph\,cm$^{-2}$\,s$^{-1}$) & $(1.24\pm0.01) \times 10^{-2}$ \\
\hline
\multirow{4}{*}{Power-law$^{b}_{\rm abs.}$} & $\Gamma$ & $l_{\rm 1}$ \\
& Norm.\,(ph\,cm$^{-2}$\,s$^{-1}$) & $(6.48^{+1.16}_{-1.10}) \times 10^{-3}$ \\
& $N_{\rm H}$\,(cm$^{-2}$) & $(1.15^{+0.30}_{-0.14}) \times 10^{23}$ \\
& log\,$\xi$ & $< 0.50$ \\
\hline
\multirow{4}{*}{\textsc{reflionx}$^{c}_{\rm abs.}$} & $\Gamma$ & $l_{\rm 1}$ \\
& $\xi$\,(erg\,cm\,s$^{-1}$) & $10_{p}$ \\
& $A_{\rm Fe}$ & $0.23\pm0.04^{^{l_{\rm 2}}}$ \\
& Norm.\,(ph\,cm$^{-2}$\,s$^{-1}$) & $(8.73^{+0.46}_{-1.00}) \times 10^{-5}$ \\
\\
\multirow{2}{*}{Absorber$^{d}$} & $N_{\rm H}$\,(cm$^{-2}$) & $(1.14^{+0.84}_{-0.65}) \times 10^{23}$ \\
& log\,$\xi$ & $< 1.12$ \\
\hline
\multirow{4}{*}{\textsc{reflionx}$^{e}_{\rm abs.}$} & $\Gamma$ & $l_{\rm 1}$ \\
& log\,$\xi$ & $3.8^{+0.2}_{-0.1}$ \\
& $A_{\rm Fe}$ & $l_{\rm 2}$ \\
& Norm.\,(ph\,cm$^{-2}$\,s$^{-1}$) & $(4.00^{+0.53}_{-0.54}) \times 10^{-7}$ \\
\\
\multirow{2}{*}{Absorber$^{d}$} & $N_{\rm H}$\,(cm$^{-2}$) & $(5.00^{+0.00\,p}_{-0.44}) \times 10^{24}$ \\
& log\,$\xi$ & $2.55\pm0.03$ \\
\hline
\end{tabular}
\end{table}

\subsection{Long-Term Spectral Variability}

In order to assess the nature of the long-term X-ray spectral variability of NGC 4051, we attempted to simultaneously model the 2008 data with the 2005 {\sl Suzaku} data over the broad-band 0.6--50.0\,keV energy range. The 2005 data have previously been described in detail by Terashima et al. (2009) where NGC 4051 was found to be in the low-flux state with a 0.5--10.0\,keV flux of $F = 1.32 \times 10^{-11}$\,erg\,cm$^{-2}$\,s$^{-1}$ compared to the much higher flux of $F = 5.03 \times 10^{-11}$\,erg\,cm$^{-2}$\,s$^{-1}$ in 2008. The spectrum was observed to be much harder in the low-flux state with a significant excess at energies \textgreater 10\,keV. Figure 13 shows the relative fluxes of the three observations. The large changes in spectral shape can be well described by a two-component model consisting of a soft variable component superimposed over a hard constant component (Miller et al. 2010). \\

\begin{figure}
\begin{center}
\rotatebox{-90}{\includegraphics[width=6cm]{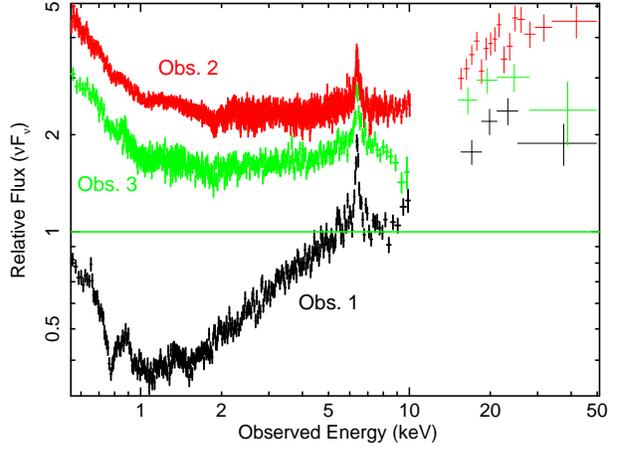}}
\end{center}
\caption{A plot showing the relative fluxes of the combined {\sl Suzaku} XIS and HXD PIN data for NGC 4051 in the observed frame. The data are unfolded against a power-law with $\Gamma = 2$. The 2005 data are shown in black. The two separate 2008 observations are shown in red and green corresponding to the 275 and 78\,ks exposures respectively.} 
\end{figure}

We simultaneously applied the best-fitting model described in Section 6.4 to the 2005 and 2008 XIS and HXD {\sl Suzaku} data. We also included the second, shorter exposure (78\,ks) observation from 2008 in our joint-analysis. We tested to see if the fully-covering warm absorber parameters varied between observations but the best-fitting values were consistent within the errors across the 2005 and 2008 data. We tied together all model parameters between the three observations except for the relative normalisations of the absorbed and intrinsic power-law components and the near-neutral and ionised reflection components (\textsc{reflionx}) which we allowed to vary to account for the long-term spectral variability. We find that the normalisations of the absorbed power-law and the near-neutral reflection component appear to be largely constant across the three observations (see Table A2), consistent with the findings of Miller et al. (2010). We note that slight changes in the flux of the highly ionised reflector can be observed between 2005 and 2008. However, we find that the spectral variability can ultimately be described by large changes in the normalisation of the intrinsic unabsorbed power-law component which increases by a factor of $\sim$7 from the low-flux state in 2005 to the high-flux state in 2008. This corresponds to a high covering fraction of the partial-coverer of $\sim$70 per cent in the 2005 low-flux data compared to a much lower covering fraction of $\sim$30 per cent when in the high-flux state and supports the notion of an empirical two-component model whereby the constant reflection component contributes to the majority of the hard X-ray flux whereas changes in the normalisation of the intrinsic power-law account for the variations in spectral shape at lower energies. The 0.5--100\,keV fluxes corresponding to the relative normalisations of the power-law and reflection components are given in Table 7. A plot of the relative contributions of the individual model components to the joint-fit is shown in Figure 14. \\

We note that upon considering the joint-fit of all {\sl Suzaku} observations, we also find that an additional emission line at $E_{\rm line} = 5.44\pm0.03$\,keV is statistically required in the 2005 data, possibly attributed to Cr\,K$\alpha$. Modelling this with a Guassian profile with an intrinsically narrow fixed width of $\sigma = 0$\,eV provides a measurement of the equivalent width of $EW = 42^{+15}_{-12}$\,eV (consistent with Turner et al. 2010) and improves the fit statistic by $\Delta \chi^{2} = 20$. We also include this line in the 2008 data and find that the flux of the line appears to be consistent with the 2005 data with the equivalent width dropping to $EW = 15^{+5}_{-8}$\,eV against the observed continuum as expected. This suggests that the emission line may have remained constant in flux over this three-year period. A detailed analysis of the nature of this line is given by Turner et al. (2010) with its possible origins discussed in a companion paper (Turner \& Miller 2010). \\

This is then considered to be our final model which appears to describe the data well with all remaining free parameters across all three observations consistent with the values listed in Table 6 resulting in a final fit statistic of $\chi^{2} / d.o.f. = 560 / 549$.

\begin{figure}
\begin{center}
\rotatebox{-90}{\includegraphics[width=6cm]{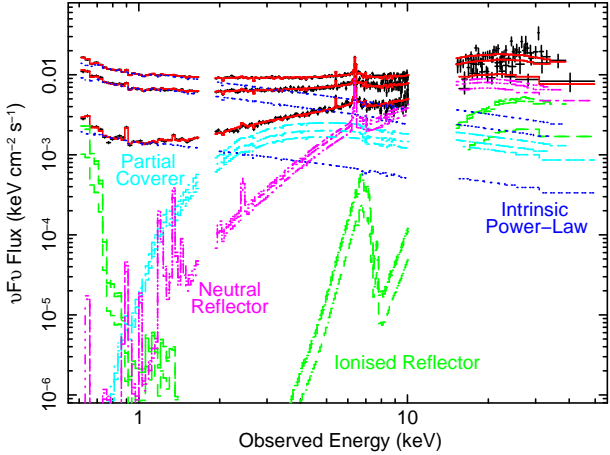}}
\end{center}
\caption{A plot showing the relative contributions of the individual model components to the three broad-band 0.6--50.0\,keV {\sl Suzaku} spectra in the observed frame. The data are shown in black with the sum of all model components superimposed in red. The absorbed and unabsorbed power-law components are shown in cyan and navy blue respectively. The absorbed neutral and ionised reflectors are shown in magenta and green respectively. Note that the absorbed power-law component and the reflection components appear to remain largely constant in flux across the three observations with the changes in spectral shape instead accounted for by significant variations in the normalisation of the intrinsic unabsorbed power-law component.}
\end{figure}

\begin{table}
\centering
\caption{The upper-half of the table shows the fluxes corresponding to the best-fitting values for the power-law (both absorbed and unabsorbed), reflected continua and Fe\,K$\alpha$ components obtained in Section 6.5. The fluxes are given across the 0.5--100\,keV energy band and have units $10^{-12}$\,erg\,cm$^{-2}$\,s$^{-1}$. The covering fraction of the partial-coverer is also shown for each observation. The lower-half of the table shows these values expressed as a ratio compared to the values obtained for the 2005 {\sl Suzaku} data. Obs.\,1 corresponds to the 2005 data whereas Obs.\,2 and 3 correspond to the two 2008 observations (275 and 78\,ks respectively). It can be seen that the fluxes of the reflection components, the absorbed power-law and the neutral Fe\,K$\alpha$ emission line are consistent with remaining largely constant across all epochs whereas large variations in the flux of the intrinsic unabsorbed power-law are apparent between observations with the flux considerably lower in 2005 when the source was in a historically low-flux state.}
\begin{tabular}{l c c c}
\hline\hline
& \multicolumn{3}{c}{Flux$_{\rm 0.5-100\,keV}$\,($\times 10^{-12}$\,erg\,cm$^{-2}$\,s$^{-1}$)} \\
Component & Obs.\,1 (2005) & Obs.\,2 (2008) & Obs.\,3 (2008) \\
\hline
Power-law & $6.63^{+0.11}_{-0.16}$ & $47.39^{+0.39}_{-0.77}$ & $31.25^{+0.26}_{-0.54}$ \\
Power-law$_{\rm abs.}$ & $6.66^{+0.17}_{-0.13}$ & $10.01^{+0.50}_{-0.47}$ & $8.03^{+0.46}_{-0.44}$ \\
\textsc{reflionx}$_{\rm neutral}$ & $17.31^{+0.54}_{-0.79}$ & $23.98^{+0.91}_{-1.76}$ & $21.25^{+1.00}_{-1.24}$ \\
\textsc{reflionx}$_{\rm ionised}$ & $3.90^{+1.07}_{-1.17}$ & $10.41^{+1.57}_{-1.52}$ & $9.23^{+2.22}_{-1.99}$ \\
Fe\,K$\alpha$ & $0.21^{+0.12}_{-0.03}$ & $0.25\pm0.04$ & $0.30^{+0.06}_{-0.05}$ \\
$f_{\rm cov}$ & $\sim$70\% & $\sim$30\% & $\sim$40\% \\
\hline
& \multicolumn{3}{c}{Ratio} \\
Component & Obs.\,1 (2005) & Obs.\,2 (2008) & Obs.\,3 (2008) \\
\hline
Power-law & 1 & $7.15\pm0.23$ & $4.71^{+0.16}_{-0.15}$ \\
Power-law$_{\rm abs.}$ & 1 & $1.50^{+0.11}_{-0.10}$ & $1.21^{+0.09}_{-0.10}$ \\
\textsc{reflionx}$_{\rm neutral}$ & 1 & $1.39^{+0.12}_{-0.15}$ & $1.23^{+0.12}_{-0.11}$ \\
\textsc{reflionx}$_{\rm ionised}$ & 1 & $2.67^{+1.72}_{-0.88}$ & $2.34^{+1.85}_{-0.88}$ \\
Fe\,K$\alpha$ & 1 & $1.19^{+0.42}_{-0.55}$ & $1.43^{+0.57}_{-0.67}$ \\
\hline
\end{tabular}
\end{table}

\section{Discussion}

\subsection{Highly Ionised Outflow}

The {\sl Suzaku} XIS spectrum of NGC 4051 reveals two absorption lines with centroid energies at $\sim$6.8 and $\sim$7.1\,keV, as described in Section 6.3. This suggests that the absorption is significantly blueshifted and could be the signature of a highly ionised, high velocity outflow. Modelling with an \textsc{xstar} grid with a turbulence velocity of $\sigma = 3\,000$\,km\,s$^{-1}$ requires one zone of absorption with a best-fitting column density of $N_{\rm H} = 8.4^{+1.9}_{-2.0} \times 10^{22}$\,cm$^{-2}$ and an ionisation parameter of log\,$\xi = 4.1^{+0.2}_{-0.1}$ (as per Section 6.3). Assuming that the lines correspond to K-shell absorption of Fe\,\textsc{xxv} and Fe\,\textsc{xxvi}, the blueshift of the zone corresponds to an outflow velocity of $v_{\rm out} = -(5\,800^{+860}_{-1200})$\,km\,s$^{-1}$, largely consistent with the findings of Pounds et al. (2004a) and Terashima et al. (2009). It is also conceivable that this is the highly ionised signature of the high velocity outflow detected by Steenbrugge et al. (2009) with the {\sl Chandra} LETG which they deduce to have a comparable outflow velocity of $v_{\rm out} \sim -4\,500$\,km\,s$^{-1}$ although with an ionisation parameter $\sim$10 times lower (i.e. log\,$\xi \sim 3.1$). \\

\subsubsection{The Kinematics of the Highly Ionised Absorption}

For a homogeneous, radial, outflowing wind, the column density along the line of sight is given by $N_{\rm H} = \int^{R_{\rm out}}_{R_{\rm in}} n_{\rm e}(R)dR$, where $R_{\rm in}$ and $R_{\rm out}$ are the inner and outer radii along the line of sight through the wind respectively. Assuming we are looking down a spherically symmetric homogeneous wind towards the inner radius, then $R_{\rm out}$ tends to $\infty$. So, combining this with the definition of the ionisation parameter (equation 1) leads to $R_{\rm in} = L_{\rm ion} / N_{\rm H} \xi$. However, if the thickness of the wind is defined as $\Delta R = R_{\rm out} - R_{\rm in}$ then the inner radius may extend inwards if the wind is clumpy (i.e. if $R >> \Delta R$). This yields:

\begin{equation}R_{\rm in} \lesssim \frac{L_{\rm ion}}{N_{\rm H} \xi}.\end{equation}

A value for the ionising luminosity can be calculated by extrapolating the total power-law continuum of the broad-band {\sl Suzaku} model from the 2008 data and integrating from 1 to 1\,000 Rydbergs. This gives a value of $L_{\rm ion} \sim 1.4 \times 10^{43}$\,erg\,s$^{-1}$. Combining this with the best-fitting values of the column density and the ionisation parameter from the \textsc{xstar} model (Table 5) then yields a value for the radius of the absorbing material of $R_{\rm in} \lesssim 1.3 \times 10^{16}$\,cm which is of the order of $\lesssim$ 0.004\,pc or $\lesssim$ 5\,l-d from the central engine. Assuming that the material escapes to infinity then a lower limit on the radius can also be calculated by considering the outflowing material's escape velocity where $v_{\rm esc} = \sqrt{\frac{2 G M}{R}}$. Rearranging for $R$ gives a lower limit of $R > 1.4 \times 10^{15}$\,cm placing the constraints on the radius of the highly ionised absorber to be $0.5 < R \lesssim 5$\,l-d from the central source and infers a constraint on the electron density of $7 \times 10^{6} \lesssim n_{\rm e} < 6 \times 10^{8}$\,cm$^{-3}$ (equation 1). The constraints on the radius derived here appear to be consistent with the radius of 0.5--1\,l-d found by Krongold et al. (2007) for their highly ionised zone of absorption found in the {\sl XMM-Newton} RGS spectrum. This is well within the 10--15\,l-d dust sublimation radius in NGC 4051 (see Krongold et al. 2004 fig.10) and so cannot have an origin in the molecular torus. Instead, as the location of the highly ionised absorber also appears to be contained within the location of the He\,\textsc{ii} broad emission line region (BELR; Krongold et al. 2007), an accretion disc wind is perhaps a more likely origin for this absorption signature. \\

By considering the case of a homogeneous, spherical flow under the assumption that the outflow velocity remains approximately constant on the compact scales observed here (although decelerating at some larger radius), an estimate on the mass outflow rate of the highly ionised zone can be calculated from the simple conservation of mass:

\begin{equation}\dot{M}_{\rm out} = 4 \pi b \frac{L_{\rm ion}}{\xi} v_{\rm out} m_{\rm p},\end{equation}

where $L_{\rm ion} / \xi = n_{\rm e} R^{2}$. Here, $\dot{M}_{\rm out}$ is the mass outflow rate in units g\,s$^{-1}$ subtending a solid angle, $4 \pi b$ (where $0 \leq b \leq 1$), where $L_{\rm ion}$ is is the ionising luminosity from 1 to 1\,000 Rydbergs in units erg\,s$^{-1}$, $\xi$ is the ionisation parameter in units erg\,cm\,s$^{-1}$, $v_{\rm out}$ is the outflow velocity in units cm\,s$^{-1}$ and $m_{\rm p}$ is the rest mass of a proton. Initially assuming that the absorber is fully covering (i.e. $b = 1$) and taking the best-fitting values of the ionisation parameter and outflow velocity given in Table 5 and the value for the ionising luminosity derived above yields an estimate on the mass outflow rate of $\dot{M}_{\rm out} \sim 0.2$\,M$_{\odot}$\,yr$^{-1}$. This value then translates into an estimate of the kinetic output of the outflow via:

\begin{equation}\dot{E}_{\rm out} = \frac{1}{2} \dot{M}_{\rm out} v^{2}_{\rm out},\end{equation} 

which corresponds to a value of $\dot{E}_{\rm out} \sim 2 \times 10^{42}$\,erg\,s$^{-1}$, which is approximately 3--4 per cent of the bolometric output of the AGN (Ogle et al. 2004). The momentum rate of the outflow is then $\dot{M}_{\rm out} v_{\rm out} \sim 8 \times 10^{33}$\,g\,cm\,s$^{-2}$. \\

For a momentum-driven outflow whereby $\tau \sim 1$ (i.e. where the photons scatter once before escaping and so all momentum is transferred to the outflow) accreting at near-Eddington (i.e. $\dot{M}_{\rm out} \sim \dot{M}_{\rm EDD}$), the total wind momentum flux must be of the same order as the photon momentum flux (King 2010), i.e.:

\begin{equation}\dot{M}_{\rm out} v_{\rm out} \approx \frac{L_{\rm EDD}}{c}.\end{equation}

Assuming a black hole mass of $M_{\rm BH} = 1.73 \times 10^{6}$\,M$_{\odot}$ (Denney et al. 2009) for NGC 4051, the Eddington luminosity is found to be $L_{\rm EDD} \sim 2.2 \times 10^{44}$\,erg\,s$^{-1}$ which yields $L_{\rm EDD} / c \sim 7.5 \times 10^{33}$\,erg\,cm$^{-1}$; a value comparable to the calculated outflow momentum rate of $\dot{M}_{\rm out} v_{\rm out} \sim 8 \times 10^{33}$\,g\,cm\,s$^{-2}$. So it can be seen that equation 6 is largely satisfied if we assume that the highly ionised outflow is momentum-driven by Thomson scattering such as in the similar case of the quasi-stellar object (QSO) PG1211+143 (King \& Pounds 2003; Pounds et al. 2003). \\

Furthermore, the Eddington luminosity is also defined by: 

\begin{equation}L_{\rm EDD} = \eta \dot{M}_{\rm EDD} c^{2},\end{equation} 

where $\eta$ is the efficiency of mass-to-energy conversion and takes a value between 0 and 1. Combining this with equation 6 yields an expression for the wind velocity (King et al. 2010):

\begin{equation}v_{\rm out} \approx \frac{\eta}{\dot{m}} c,\end{equation}

where $\dot{m} = \dot{M}_{\rm acc} / \dot{M}_{\rm EDD}$ is the accretion rate in Eddington units. Since $v_{\rm out} = 0.02c$ here, a value of $\eta \sim 0.02$ would be required to sustain the Eddington luminosity of this source if it were accreting at the Eddington rate (i.e. $\dot{m} \sim 1$). Note that the maximum value of $\eta$ is 0.06 for a non-rotating, Schwarzschild black hole. \\

The bolometric luminosity of NGC 4051 calculated by Ogle et al. (2004) and corrected for the distance of 15.2\,Mpc assumed here is $L_{\rm bol} \sim 7 \times 10^{43}$\,erg\,s$^{-1}$. Comparing this to the Eddington luminosity suggests that this object has an accretion rate of only $\sim$30 per cent of $\dot{M}_{\rm EDD}$ which would correspond to an observed mass accretion rate of $\dot{M}_{\rm acc} = L_{\rm bol} / \eta c^{2} \sim 0.06$\,M$_{\odot}$\,yr$^{-1}$ assuming an efficiency of $\eta = 0.02$. However, the observed luminosity of the system with respect to Eddington may be underestimated if some of the emission is obscured by a Compton-thick wind and scattered out of the line of sight (a notion perhaps supported by the apparent presence of high column densities towards this source). So it is perhaps conceivable that NGC 4051 is accreting at an appreciable fraction of the Eddington rate. Furthermore, Sim et al. (2010) show that radiatively-driven accretion disc winds can exist with similar properties to those that we observe here for sources with luminosities of a few tens of per cent of $L_{\rm EDD}$.

\subsubsection{The Covering Fraction}

From the consideration of mass conservation (equation 4), it can be seen that the mass outflow rate is dependent upon the covering factor, $b$, which is likely to be some significant fraction of $4 \pi$\,sr. The calculations in Section 7.1.1 were based on the assumption that the outflow is a radial flow in a spherical geometry with $b = 1$. However, if the outflow takes the form of a bi-conical geometry (Elvis 2000), as proposed by Krongold et al. (2007) for this source then the covering fraction may be lower. \\

The {\sl Chandra} HETG and {\sl Suzaku} XIS spectra reveal the presence of two highly ionised Fe\,K emission lines most likely originating from Fe\,\textsc{xxv} and Fe\,\textsc{xxvi}. If these lines are the signature of the same ions associated with the highly ionised, high velocity outflow, we can attempt to obtain an estimate on the covering fraction of the material. Bianchi \& Matt (2002) calculate the predicted equivalent widths of the Fe\,\textsc{xxv} and Fe\,\textsc{xxvi} emission lines against the total (primary + reprocessed) continuum for a given photon index and column density. Figures 8 and 9 of Bianchi \& Matt (2002) show that for a column density of $N_{\rm H} = 10^{23}$\,cm$^{-2}$ and a photon index of $\Gamma = 2.4$, the expected equivalent widths are $\sim$50\,eV and $\sim$15\,eV for the Fe\,\textsc{xxv} and Fe\,\textsc{xxvi} emission lines respectively. As the column density of the highly ionised absorption that we observe here with {\sl Suzaku} is $N_{\rm H} \sim 8 \times 10^{22}$\,cm$^{-2}$ (see Section 6.3) and the best-fitting photon index is $\Gamma \sim 2.5$ (see Section 6.4) in the 2008 data, we can make a direct comparison. Within the errors, the equivalent widths of $\sim$25\,eV and $\sim$10\,eV obtained for the Fe\,\textsc{xxv} and Fe\,\textsc{xxvi} emission in the {\sl Suzaku} data respectively (see Table 5) suggest that the values are on the order of $\sim$50 per cent lower than the predicted maximum values perhaps implying that the covering factor is on the order of $b \sim 0.5$. \\

Furthermore, Tombesi et al. (2010) analysed a sample of 42 radio-quiet AGN observed with {\sl XMM-Newton}. They detected 22 absorption lines at rest-frame energies \textgreater 7.1\,keV, implying that high velocity outflows may be a common phenomenon in radio-quiet AGN. The fraction of AGN having at least one feature with a blue-shifted velocity in their sample is $17 / 42$, corresponding to $\sim$40 per cent of the objects. They also find that the global covering fraction of the absorbers that they detect is estimated to be in the range $C \sim$ 0.4--0.6, suggesting that the outflowing winds generally have large opening angles. \\

We can also obtain a rough estimate on $b$ by comparing the observed accretion rate of the source ($\dot{M}_{\rm acc} \sim 0.06$\,M$_{\odot}$\,yr$^{-1}$; see Section 7.1.1) with the calculated spherical outflow rate from equation 4 ($\dot{M}_{\rm out} \sim 0.2$\,M$_{\odot}$\,yr$^{-1}$) under the assumption that the mass outflow rate either does either not exceed or is of the same order as the mass accretion rate (i.e. $\dot{M}_{\rm out} \lesssim \dot{M}_{\rm acc}$). This suggests a covering fraction of $b \lesssim 0.3$ although this could be higher if the observed luminosity with respect to Eddington is underestimated somewhat perhaps due to partially-covering Compton-thick clouds in our line of sight. So it seems that although the absorption component is unlikely to be fully covering, the observational evidence suggests that the outflow is not highly collimated with $b$ still likely to correspond to a significant fraction of 1 which appears to be consistent with the detailed reverberation analysis of Miller et al. (2010) who state that a global covering fraction of $\gtrsim 40$ per cent is required to match the observations.

\subsubsection{Implications for Feedback}

Given the high accretion rates required for the growth of supermassive black holes, significant outflows may be a consequence of near-Eddington accretion (King 2010). If so, they could be an important factor in galaxy evolution and in establishing the observed $M$--$\sigma$ and $M$--$M_{\rm bulge}$ relations between the supermassive black hole and its host galaxy (Magorrian et al. 1998; Ferrarese \& Merritt 2000; Gebhardt 2000). \\

Taking the bolometric luminosity of $L_{\rm bol} \sim 7 \times 10^{43}$\,erg\,s$^{-1}$ of NGC 4051, we can calculate an estimate for the mass accretion rate of $\dot{M}_{\rm acc} \sim 0.06$\,M$_{\odot}$\,yr$^{-1}$. Thus, for the black hole to accrete a mass of $M_{\rm BH} = 1.73 \times 10^{6}$\,M$_{\odot}$ (Denney et al. 2009), the Salpeter e-folding time would be $\sim$$3 \times 10^{7}$\,yr. As calculated in Section 7.1.1, the energy output of the high velocity outflow is $\dot{E}_{\rm out} \sim 2 \times 10^{42}$\,erg\,s$^{-1}$ assuming that $b \sim 1$. Integrating this over the e-folding time of the AGN gives a total energy output of $E_{\rm tot} \sim 10^{57}$\,erg (even if we assume a conservative value of $b = 0.1$, this still provides a total output of $E_{\rm tot} \sim 10^{56}$\,erg). This value can then be compared with the binding energy of the bulge of the host galaxy. Marconi \& Hunt (2003) and H\"{a}ring \& Rix (2004) provide an observable relationship between the mass of the central object and the mass of the galaxy bulge, i.e. $M_{\rm BH} \sim 10^{-3} M_{\rm bulge}$. This can then be used to calculate the binding energy of NGC 4051 whereby $B.E. \sim \sigma^{2} M_{\rm bulge}$. Nelson \& Whittle (1995) provide a value for the velocity dispersion of NGC 4051 of $\sigma = 88$\,km\,s$^{-1}$ which then yields an estimate on the binding energy of $B.E. \sim 3 \times 10^{56}$\,erg, comparable with the likely total energy deposited by the wind (assuming that the wind persists for the lifetime of the AGN). Therefore, it can be seen that the total energy output of the high velocity outflow could have a considerable influence on its host galaxy environment. Indeed it may be conceivable that NGC 4051 is a low mass analague of the QSOs PG1211+143 (Pounds \& Reeves 2009) and PDS 456 (Reeves et al. 2009) which may deposit as much as $10^{60}$\,erg throughout the lifetime of the AGN.

\subsection{The Soft X-ray Absorption / Emission}

The HETG spectrum shows evidence for a wealth of absorption lines in the soft X-ray band. Modelling this warm absorber with \textsc{xstar} requires four distinct ionisation zones of gas with the ionisation parameter ranging from log\,$\xi = -0.86$ to log\,$\xi = 2.97$ and column densities on the order of $10^{20}$--$10^{21}$\,cm$^{-2}$ (see Table 3). The zones appear to be outflowing with velocities on the order of a few 100\,km\,s$^{-1}$ with the general trend appearing to be that the larger outflow velocities appear to correspond to more highly ionised zones. This is perhaps suggestive of a geometry whereby all of the zones form part of the same extended outflowing wind with the higher velocity, more highly ionised components originating much closer to the central source. Indeed, as the mass outflow rate (equation 4) and the ionisation parameter (equation 1) combine to give:

\begin{equation}\dot{M}_{\rm out} \propto \frac{L_{\rm ion} v_{\rm out}}{\xi} = const.\end{equation}

(King et al. 2010), such a trend between ionisation parameter and outflow velocity would be expected to be observed if the mass outflow rate is to be conserved. Such a correlation has previously been noted in NGC 4051 by Pounds et al. (2004a) and, if confirmed, could provide strong evidence of a cooling shock (King et al. 2010). \\

Using equation 3, we are able to obtain upper limits for the radius of the absorbing zones of gas (see Table 8). However, these values are largely unconstrained and result in very conservative estimates on the order of kpc for the two most lowly ionised zones. The constraints on the medium and higher ionisation zones are a little tighter with the radius of the material falling within $\sim$70 and $\sim$2\,pc from the central source respectively. We note that the absorbers could exist at smaller radii still if the material is somewhat clumpy or filamentary. \\

Calculations of the mass outflow rate of zones 1--4 using equation 4 prove to significantly exceed the Eddington accretion rate of the source by several orders of magnitude since they assume that the absorber forms part of a fully-covering, homogeneous, radial flow. However, if we assume that the mass outflow rate does not exceed the observed mass accretion rate since the energy required to accelerate the outflow must come from the infalling gas (i.e. $\dot{M}_{\rm out} \lesssim 0.06$\,M$_{\odot}$\,yr$^{-1}$), we can scale the soft zones accordingly and place upper limits on the covering fraction. This results in values ranging from $b \lesssim 4.4 \times 10^{-4}$ to $b \lesssim 0.56$ for the lowest and highest ionisation zones in the HETG data respectively. The values for the lower ionisation zones are generally consistent with the opening angles of the warm absorber components detected by Steenbrugge et al. (2009) with the {\sl Chandra} LETG. We also calculated upper limits for the kinetic output of the outflow using our assumption about the mass outflow rate stated above and equation 5. This resulted in values ranging from $\dot{E}_{\rm out} \lesssim 1.8 \times 10^{39}$\,erg\,s$^{-1}$ to $\dot{E}_{\rm out} \lesssim 2.9 \times 10^{40}$\,erg\,s$^{-1}$ for the five zones of absorption, i.e. very small fractions of the total bolometric output of the AGN in constrast to the high velocity outflow that we observe with {\sl Suzaku} which has a kinetic output at least two orders of magnitude higher and could potentially be highly significant in terms of galactic feedback. These values are summarised in Table 8. \\

\begin{table}
\centering
\caption{Table showing the upper limits of the radius, $R$, kinetic energy output, $E_{\rm K}$, and covering fraction, $b$, obtained for the five individual zones of absorption detected in the {\sl Chandra} HETG data (see Section 5.1) assuming that the mass outflow rates for each zone does not exceed or is on the order of $\sim$0.1\,M$_{\odot}$\,yr$^{-1}$. The absorption zones directly correspond to those listed in Table 4. We also include the highly ionised zone of absorption detected with {\sl Suzaku} for comparison.}
\begin{tabular}{c c c c}
\hline\hline
Absorption & $R$ & $E_{\rm K}$ & $b$ \\
Component & (pc) & (erg\,s$^{-1}$) & \\
\hline
Zone 1 & $\lesssim 8.7 \times 10^{4}$ & $\lesssim 1.8 \times 10^{39}$ & $\lesssim 4.4 \times 10^{-4}$ \\
Zone 2 & $\lesssim 1.2 \times 10^{4}$ & $\lesssim 7.4 \times 10^{39}$ & $\lesssim 5.2 \times 10^{-3}$ \\
Zone 3a & $\lesssim 75$ & $\lesssim 1.7 \times 10^{40}$ & $\lesssim 0.10$ \\
Zone 3b & $\lesssim 65$ & $\lesssim 3.8 \times 10^{40}$ & $\lesssim 0.05$ \\
Zone 4 & $\lesssim 1.8$ & $\lesssim 2.9 \times 10^{40}$ & $\lesssim 0.56$ \\
& & & \\
High $\xi$ & $0.0004 < R \lesssim 0.004$ & $\sim$$2 \times 10^{42}$ & $\sim$0.5 \\
\hline
\end{tabular}
\end{table}

We also detect several narrow emission features at energies \textless 2\,keV. The ionised reflector is unable to model all of the emission lines and, in particular, cannot account for the forbidden transitions from O\,\textsc{vii}, Ne\,\textsc{ix} and Si\,\textsc{xiii}. Previous studies with the RGS on-board {\sl XMM-Newton} have claimed the presence of RRC (Ogle et al. 2004; Pounds et al. 2004a) associated with H-like and He-like ions of elements such as O and Ne, especially when the source is found to be in an extended period of low flux, suggesting that the soft X-ray emission lines may have a photo-ionised origin. Indeed an origin in photo-ionised gas (either formed in the NLR or in an extended photo-ionised outflow) has been claimed for soft X-ray lines in many other objects (e.g. Bianchi et al. 2006). If the emission lines detected here do indeed originate in a photo-ionised plasma, an estimate on the electron density of the O\,\textsc{vii} emission (the strongest soft X-ray emission line detected here) can be calculated using the $R$ ratio for He-like ions of Porquet \& Dubau (2000), which is defined as $R (n_{\rm e}) = \frac{z}{x + y}$ where $z$ is the intensity of the forbidden line and $x$ and $y$ are the intensities of the intercombination lines. An upper limit on the flux of the O\,\textsc{vii} intercombination line(s) can be found by including an additional Gaussian in the HETG spectrum with the centroid energy fixed at 568.5\,eV and the intrinsic width of the line tied to that of the associated forbidden transition. This component is unrequired by the data and provides an upper limit on the line flux of $F_{\rm line} < 2.85 \times 10^{-5}$\,photons\,cm$^{-2}$\,s$^{-1}$. Combining this with the flux of the O\,\textsc{vii} forbidden line of $F_{\rm line} = 1.25^{+0.37}_{-0.32} \times 10^{-4}$\,photons\,cm\,s$^{-1}$ (see Table A1) provides a lower limit on the value of the $R$ ratio\footnote{One caveat to this is that densities obtained with the $R$ ratio are also dependant upon photoexcitation of the associated resonance line and optical depth effects.} of $R > 3$. Then, by considering the relationship between the $R$ ratio and electron density for O\,\textsc{vii} shown in fig.8 of Porquet \& Dubau (2000), an upper limit on the electron density for the O\,\textsc{vii} emission can be obtained which is on the order of $n_{\rm e} < 10^{11}$\,cm$^{-3}$. If the majority of O\,\textsc{vii} absorption is being modelled by zone 2 (see Table 3 and Figure 7), then by assuming that the emission has an ionisation parameter of log\,$\xi \sim 0.6$, equation 1 allows us to place a lower limit on the radius of this emission of $R > 3.8 \times 10^{15}$\,cm, which is on the order of $\sim$$10^{4}$\,$R_{\rm g}$ or $\sim$1.5\,l-d. This radius is largely consistent with the value of $R < 100$\,l-d found by Steenbrugge et al. (2009) and the value of $R < 3.5$\,l-d for the lowly ionised component detected by Krongold et al. (2007).

\subsection{The Nature of the Long-Term Spectral Variability}

NGC 4051 is found to be highly variable on both short- (ks) and long-term (years) time-scales with the spectral trend being that the X-ray spectrum flattens above a few keV as the source flux drops. Occasionally NGC 4051 is found to fall into an extended period of low flux (e.g. Uttley et al. 1999) revealing a very hard X-ray spectrum. The general consensus is that the primary power-law disappears from view as the flux drops leaving the remaining emission dominated by a hard reflection component (e.g. Guainazzi et al. 1998; Pounds et al. 2004a). Such a low-flux state was observed in 2005 by Terashima et al. (2009) when a very strong hard excess was observed at energies \textgreater 10\,keV which, if modelled with neutral reflection (i.e. a \textsc{pexrav} component), returns a value of $R \sim 7$ (where $R = 1$ corresponds to reflection from a semi-infinite slab). This is also apparent in Table 7 where it can be seen that the 0.5--100\,keV flux of the reflector is significantly higher than that of the intrinsic power-law in the 2005 data. However, the excess could not be accounted for by reflection alone due to the relatively weak Fe\,K$\alpha$ emission component ($EW \sim 140$\,eV). Instead, Terashima et al. (2009) were able to model the broad-band spectra by introducing a partially-covered power-law component which gives rise to the hard spectral shape. \\ 

In Section 6.4, we successfully modelled the broad-band 2008 {\sl Suzaku} data with a model consisting of an intrinsic power-law component, a partial-coverer, near-neutral and ionised reflection, ionised absorption from a variety of ionisation states and several soft X-ray photo-ionised emission lines. This model accounts for the spectral curvature well with the partial-coverer appearing to have a low covering fraction of $\sim$30 per cent. We then attempted to simultaneously fit all three broad-band {\sl Suzaku} spectra from 2005 and 2008 using the same model (see Section 6.5). We find that the long-term spectral variability can be accounted for largely by changes in the normalisation of the intrinsic (unabsorbed) power-law. Since the absorbed power-law remains unchanged, the spectral variability cannot be accounted for by simple changes in the covering fraction of the partial-coverer alone.  If this were the case, then the total flux (intrinsic + absorbed) would not vary. However, since direct changes in the covering fraction are observed between observations, the spectral variability cannot be accounted for by intrinsic changes in the luminosity of the source alone either. Therefore, either the covering fraction of the Compton-thin partial-coverer and the luminosity of the source are inversely correlated or the changes in $L$ are not intrinsic. \\ 

If the covering fraction and luminosity are inversely correlated, one possible cause of this could be that the hot corona above the disc may be varying in area. This could have the effect of concentrating the continuum emission from a smaller region, therefore increasing the proportion which is absorbed by the partial-coverer and lowering the luminosity, such as is observed in the 2005 data. Alternatively, the accretion disc itself may become truncated which could also have the effect of lowering the intrinsic luminosity and concentrating the continuum emission over a smaller area. Taking values of $\dot{M} = 0.06$\,M$_{\rm \odot}$\,yr$^{-1}$, $M_{\rm BH} = 1.73 \times 10^{6}$\,M$_{\rm \odot}$ and $\alpha \sim 1$, from the Shakura \& Sunyaev (1973) solution for a thin accretion disc, an estimate can be made on the viscous timescale at the radius of the innermost stable circular orbit for a Schwarzschild black hole (i.e. 6\,$R_{\rm g}$) for NGC 4051 of $t_{\rm visc} \sim 10^{7}$\,s (Frank, King \& Raine 2002). This suggests that the disc could vary on timescales as short as a few months or years and so could be a possible cause of the long-term variations which we observe. A third possibility remains that if the intrinsic ionising luminosity increases, the absorbing clouds may become ionised and so more transparent to the continuum emission. This could result in an apparent change in the covering fraction of the absorber versus luminosity and therefore lead to the observed spectral changes between 2005 and 2008. \\

Alternatively, if the observed changes in $L$ are not intrinsic, then another possible reason for the observed changes in the flux of the intrinsic power law could be due to the intrinsic continuum emission disappearing from view as the covering fraction of a further Compton-thick, variable partial-coverer increases therefore resulting in an apparent drop in continuum flux. Thus, this leaves behind a constant, hard, reflection component which then dominates the spectrum at low fluxes leading to the spectral trend observed by Guainazzi et al. (1998) and Pounds et al. (2004a). If the proposed Compton-thick clouds form part of the same system as the more distant Compton-thin clouds, this could explain the correlation with the line-of-sight covering fraction of the Compton-thin partial-coverer (i.e. as $f_{\rm cov}$ of the Compton-thick clouds increases, $f_{\rm cov}$ of the Compton-thin clouds increases too). Such an interpretation of accounting for spectral variability by allowing for changes in the covering fraction of an absorbing layer of gas has also been used as a solution to several other well-studied AGN (e.g. 1H 0419-577, Pounds et al. 2004b; NGC 1365, Risaliti et al. 2007; MCG-6-30-15, Miller, Turner \& Reeves 2008; NGC 3516, Turner et al. 2008). Then, if the origin of the distant, near-neutral reflector that we observe is located outside of the partial-coverer (for instance, the molecular torus), the apparent value of $R > 1$ observed in the 2005 data could simply be due to a delay in the reflection component responding to a previous apparent brighter state since the continuum emission from the central engine may also be obscured by the Compton-thick clouds at the distance of the torus. \\

These results are consistent with the findings of Miller et al. (2010) who performed PCA on the three {\sl Suzaku} observations of NGC 4051 from 2005 and 2008. They were able to decompose the spectra into the principle modes of variation and confirmed that a constant, hard component is present in the spectra at all times with a highly variable soft X-ray component superimposed. This was attributed to variations in the covering fraction of a layer of absorbing gas with a covering fraction of $\gtrsim 40$ per cent obscuring the central source on long time-scales.  They suggested that the strong variability on short (ks) time-scales however, is intrinsic to the source. \\

The flux of the narrow, unresolved Fe\,K$\alpha$ component appears to remain constant between the 2005 and 2008 {\sl Suzaku} observations (see Table 7) with the equivalent width of the line decreasing against the observed continuum as the flux level of the continuum increases (see Table A2). The constancy of the flux over long-term time-scales suggests that near-neutral Fe\,K$\alpha$ emission does therefore not respond to apparent changes in the continuum level and is perhaps supportive of an origin in very distant material. The constant flux of the neutral Fe\,K$\alpha$ emission line is also observed in the PCA of Miller et al. (2010; see figure 2) despite large apparent changes in the continuum flux. \\

In addition to the near-neutral Fe\,K$\alpha$ emission from distant material, we also find evidence for a weak, broader component to the K$\alpha$ emission in the HETG spectrum, as described in Section 5.2. Taking the FWHM of $16\,000^{+7\,000}_{-4\,000}$\,km\,s$^{-1}$ (i.e. $\sigma \sim 8\,000$\,km\,s$^{-1}$) of this broad Fe\,K$\alpha$ component from the HETG data allows an estimate on the radius from the central engine of the emitting material to be calculated assuming simple Keplerian motion. This yields an estimate on the radius of $R \sim 1-2 \times 10^{3}$\,$R_{\rm g}$ where $R_{\rm g} = GM / c^{2}$. Taking the value of $1.73 \times 10^{6}$\,M$_{\odot}$ for the mass of the black hole calculated by Denney et al. (2009), this corresponds to a value of $R \sim 0.1$\,l-d from the central source. Again, this is largely consistent with the findings of Miller et al. (2010) who find evidence of a weak variable hard excess at energies \textgreater 20\,keV with an associated weak, moderately broadened Fe\,K$\alpha$ component which appears to respond to changes in the continuum on a 20\,ks time-scale. This may be related to the ionised reflector which is seen to vary somewhat in the {\sl Suzaku} data (see Table 7) and may be associated with reflection occuring off the outer disc or off a disc wind similar to that calculated by Sim et al. (2010). Miller et al. (2010) also find evidence for time-lags between the hard and soft bands which are well described by reverberation from material a few light-hours away from the illuminating source. This confirms the idea that NGC 4051 has a substantial amount of optically-thick material within a few 100\,$R_{\rm g}$ from the central source and is again supportive of a partial-covering solution to account for the long-term spectral variability. \\

Finally, we also note that whether parameterising the broad-band {\sl Suzaku} data with a \textsc{pexrav} or \textsc{reflionx} model, the data seem to prefer a sub-Solar abundance of iron with a best-fitting value of $A_{\rm Fe} \sim 0.3$ times Solar. Aside from an intrinsically sub-Solar abundance of iron in NGC 4051, one other possible interpretation could be that this is perhaps indicative of spallation (Skibo 1997); a phenomenon whereby Fe nuclei are fragmented into lighter nuclei by the impact of high energy cosmic rays, thus enhancing the abundances of lower $Z$ atoms such as Cr and Mn. Indeed, in Section 6.5 we note that we detect an additional emission line at $E_{\rm line} = 5.44$\,keV, whose centroid energy appears to coincide with the expected line energy of the neutral Cr\,K$\alpha$ transition thus suggesting an enhanced abundance of Cr. The presence of this line is discussed in a companion paper (Turner et al. 2010) where they show that the observed line at $\sim$5.44\,keV cannot be caused by a simple statistical fluctuation. Furthermore, the possible origin of this line is discussed further by Turner \& Miller (2010) where they consider an origin in a transient hotspot but ultimately favour a spallation interpretation.

\section{Conclusions}

Through a detailed analysis of the X-ray spectrum of NGC 4051 with the {\sl Chandra} HETG and the {\sl Suzaku} XIS and HXD instruments, we are able to fully parameterise the warm absorber finding zones of absorption ranging in ionisation parameter from log\,$\xi = -0.86$ to log\,$\xi = 4.1$ and ranging in column density from $N_{\rm H} \sim 10^{20}$\,cm$^{-2}$ to $N_{\rm H} \sim 10^{23}$\,cm$^{-2}$.  
The soft X-ray absorber zones appear to be outflowing with velocities on the order of a few 100\,km\,s$^{-1}$ with the general trend being that the outflow velocity increases with ionisation parameter. The kinetic output of the zones then appear to be on the order of $\dot{E}_{\rm out} \lesssim 10^{39}$--$10^{40}$\,erg\,s$^{-1}$; values which correspond to very small fractions of the total bolometric output of the AGN and are negligible with respect to the binding energy of the galactic bulge (although such kinetic luminosities could still significantly influence the ISM; Hopkins \& Elvis 2010). \\

Regarding the 2008 {\sl Suzaku} spectrum of NGC 4051, we detect the presence of two statistically significant absorption lines at $\sim$6.8 and $\sim$7.1\,keV. If these lines are the signature of He-like and H-like Fe respectively then their blueshift suggests that they are the ionised signature of the high velocity outflowing wind previously detected by Pounds et al. (2004a) and Terashima et al. (2009). We are able to model the absorption lines with a single absorber with a turbulence velocity of $\sigma = 3\,000$\,km\,s$^{-1}$ and best-fitting parameters of log\,$\xi \sim 4.1$ and $N_{\rm H} \sim 8.4 \times 10^{22}$\,cm$^{-2}$. The zone appears to correspond to an outflow velocity of $v_{\rm out} \sim -5\,800$\,km\,s$^{-1}$ ($\sim$$-0.02c$) and is perhaps the highly ionised signature of the high velocity outflowing zones of gas detected by Collinge et al. (2001) and Steenbrugge et al. (2009). 
We constrain the location of the material to be on the order of a few l-d from the black hole, well within the dusty torus for this object and may originate in an accretion disc wind perhaps depositing up to $10^{56}$--$10^{57}$\,erg throughout the lifetime of the AGN. We note that we do not, however, find any requirement in these data for the higher velocity outflow ($v_{\rm out} \sim -30\,000$\,km\,s$^{-1}$) at Fe\,K reported by Pounds \& Vaughan (2010) in their results of a recent {\sl XMM-Newton} RGS observation. \\

Finally, the long-term spectral variability can be modelled simply by allowing for changes in the normalisation of the intrinsic, unabsorbed power-law component with respect to a quasi-constant hard reflection component. This can be interpreted as being due to changes in the covering fraction of a Compton-thick absorber obscuring the intrinsic continuum emission. This is consistent with the findings of Miller et al. (2010) who, through studying the effects of reverberation in the hard band, find that a global covering factor of $\gtrsim 40$ per cent of reflecting material is required in this source. 
The constancy of the flux of the near-neutral Fe\,K$\alpha$ component across epochs suggests that it does not respond to changes in the overall spectral shape and this is also perhaps supportive of a partial-covering scenario as opposed to intrinsic changes in the continuum flux.

\section{Acknowledgements}

This research has made use of data obtained from the {\sl Suzaku} satellite, a collaborative mission between the space agencies of Japan (JAXA) and the USA (NASA). TJT would like to acknowledge the NASA/{\sl Chandra} grant GO9-0123X and NASA grant NNX 09AO92G. This research has also made use of data obtained from the {\sl Chandra} X-ray Observatory and software provided by the {\sl Chandra} X-ray Center (CXC) in the application package CIAO. We would also like to acknowledge data obtained from the High Energy Astrophysics Science Archive Research Center (HEASARC), provided by NASA's Goddard Space Flight Center.

\appendix

\section{}

\begin{table*}
\centering
\caption{Table showing the best-fitting parameters of the absorption and emission lines which we detect in the HETG spectrum \textless 2\,keV. We note that several of the absorption lines appear to be spectrally resolved with FWHM corresponding to a few hundred to $\sim$1\,000\,km\,s$^{-1}$. $\Delta C$ corresponds to the change in the $C$-statistic upon modelling the feature in the data. All parameters are given in the rest frame of the host galaxy. See Sections 4.1 and 4.2 for further details.}
\begin{tabular}{l c c c c c c c}
\hline\hline
\multicolumn{1}{c}{Line Energy} & Line Width & FWHM & EW & Line Flux & Transition $\&$ Rest-Frame Energy & Velocity Shift & $\Delta C$ \\
\multicolumn{1}{c}{(eV)} & (eV) & (km\,s$^{-1}$) & (eV) & ($\times\,10^{-6}$\,photons\,cm$^{-2}$\,s$^{-1}$) & (eV) & (km\,s$^{-1}$) & \\
\hline
555.17$^{+0.03}_{-0.01}$ & $< 0.16$ & $< 190$ & $-$(0.95$^{+0.17}_{-0.08}$) & $-$(49.3$^{+8.6}_{-4.3}$) & O\,\textsc{vi} $1s$--$2p$ (554.25) & $-$(500$^{+20}_{-10}$) & 33.3 \\
561.42$^{+0.10}_{-0.09}$ & $< 0.34$ & $< 290$ & $2.38^{+0.70}_{-0.60}$ & 125.4$^{+37.0}_{-31.9}$ & [O\,\textsc{vii}] $1s$--$2s$ (560.98) & $-$$(230 \pm 50)$ & 88.2 \\
563.87$^{+0.14}_{-0.19}$ & 0.17$^{+0.17}_{-0.16}$ & 310$^{+210}_{-200}$ & $-$(0.71$^{+0.16}_{-0.17}$) & $-$(34.5$^{+8.0}_{-8.4}$) & O\,\textsc{vi} $1s$--$2p$ (563.05) & $-$(440$^{+70}_{-100}$) & 17.2 \\
575.00$^{+0.16}_{-0.17}$ & 0.50$^{+0.19}_{-0.05}$ & 610$^{+230}_{-57}$ & $-$(1.26$^{+0.32}_{-0.10}$) & $-$(65.6$^{+16.5}_{-5.3}$) & O\,\textsc{vii} $1s$--$2p$ (573.95) & $-$(550$^{+80}_{-90}$) & 43.4 \\
652.09$^{+0.18}_{-0.22}$ & $< 0.38$ & $< 400$ & 0.97$^{+0.43}_{-0.33}$ & 33.2$^{+14.9}_{-11.2}$ & O\,\textsc{viii} $1s$--$2p$ (653.49) & $650^{+100}_{-80}$ & 30.1 \\
653.62$^{+0.14}_{-0.19}$ & 0.54$^{+0.21}_{-0.11}$ & 570$^{+140}_{-110}$ & $-$(1.41$^{+0.30}_{-0.21}$) & $-$(50.2$^{+10.8}_{-7.6}$) & O\,\textsc{viii} $1s$--$2p$ (653.49) & $-$($740^{+70}_{-90}$) & 69.1 \\
666.84$\pm0.24$ & 0.68$^{+0.27}_{-0.22}$ & 700$^{+280}_{-230}$ & $-$(1.30$^{+0.31}_{-0.43}$) & $-$(43.8$^{+10.6}_{-14.7}$) & O\,\textsc{vii} $1s$--$3p$ (665.62) & $-$($550 \pm 110$) & 48.7 \\
698.63$^{+0.16}_{-0.13}$ & $< 0.50$ & $< 490$ & $-$(0.57$^{+0.18}_{-0.16}$) & $-$(16.6$^{+5.3}_{-4.7}$) & O\,\textsc{vii} $1s$--$4p$ (697.80) & $-$($360^{+70}_{-50}$) & 19.0 \\
776.26$^{+0.19}_{-0.19}$ & 0.37$^{+0.24}_{-0.36}$ & $< 530$ & $-$(0.88$^{+0.23}_{-0.25}$) & $-$(19.7$^{+5.2}_{-5.5}$) & O\,\textsc{viii} $1s$--$3p$ (774.58) & $-$($650^{+70}_{-80}$) & 38.5 \\
827.49$^{+0.18}_{-0.25}$ & $< 0.71$ & $< 360$ & $-$(0.88$^{+0.25}_{-0.35}$) & $-$(15.9$^{+4.5}_{-6.3}$) & Fe\,\textsc{xvii} $2p$--$3d$ (825.73) & $-$(640$^{+70}_{-90}$) & 36.0 \\
874.45$^{+0.58}_{-0.85}$ & 1.13$^{+1.09}_{-0.75}$ & 890$^{+860}_{-590}$ & $-$(1.37$^{+0.58}_{-0.70}$) & $-$(22.0$^{+9.3}_{-11.2}$) & Fe\,\textsc{xviii} $2p$--$3d$ (872.65) & $-$(620$^{+200}_{-290}$) & 23.7 \\
885.14$^{+0.72}_{-0.35}$ & 1.28$^{+0.59}_{-0.50}$ & 1000$^{+460}_{-390}$ & $-$($1.2 \pm 0.5$) & $-$($18.7 \pm 7.0$) & Unidentified & & 20.5 \\
897.89$^{+0.14}_{-0.28}$ & $< 0.55$ & $< 420$ & $-$(0.91$^{+0.20}_{-0.31}$) & $-$(13.2$^{+2.8}_{-4.6}$) & Ne\,\textsc{viii} $1s$--$2p$ (895.89) & $-$(670$^{+50}_{-90}$) & 41.0 \\
906.07$^{+0.38}_{-0.42}$ & 0.72$^{+0.49}_{-0.66}$ & 550$^{+370}_{-500}$ & 1.39$^{+0.52}_{-0.48}$ & 19.5$^{+7.3}_{-6.7}$ & [Ne\,\textsc{ix}] $1s$--$2s$ (905.08) & $-$($330^{+130}_{-140}$) & 35.5 \\
915.19$^{+0.60}_{-0.83}$ & $< 1.72$ & $< 1300$ & 1.12$^{+0.55}_{-0.53}$ & 15.3$^{+7.5}_{-7.3}$ & Ne\,\textsc{ix} $1s$--$2p$ (914.80) & $-$(130$^{+200}_{-270}$) & 21.1 \\
919.72$^{+0.26}_{-0.29}$ & 0.58$^{+0.43}_{-0.53}$ & 440$^{+310}_{-400}$ & $-$(1.15$^{+0.44}_{-0.34}$) & $-$(15.8$^{+6.1}_{-4.6}$) & Fe\,\textsc{xix} $2p$--$3d$ (918.10) & $-$($530^{+90}_{-100}$) & 41.8 \\
924.01$^{+0.28}_{-0.25}$ & 0.85$\pm0.29$ & 640$^{+220}_{-210}$ & $-$(1.67$^{+0.35}_{-0.36}$) & $-$(22.6$^{+4.7}_{-4.8}$) & Ne\,\textsc{ix} $1s$--$2p$ (922.02) & $-$($650^{+90}_{-80}$) & 72.1 \\
968.29$^{+0.54}_{-0.22}$ & 0.78$^{+0.35}_{-0.53}$ & 550$^{+250}_{-390}$ & $-$(1.09$^{+0.38}_{-0.35}$) & $-$(12.9$^{+4.5}_{-4.1}$) & Fe\,\textsc{xx} $2p$--$3d$ (967.33) & $-$(300$^{+170}_{-70}$) & 38.1 \\
1011.91$^{+0.14}_{-0.19}$ & $< 0.36$ & $< 250$ & $-$(0.81$^{+0.08}_{-0.19}$) & $-$(8.7$^{+0.9}_{-2.1}$) & Fe\,\textsc{xxi} $2p$--$3d$ (1009.23) & $-$(790$^{+40}_{-60}$) & 46.7 \\
1022.48$^{+0.87}_{-0.30}$ & $< 1.14$ & $< 770$ & 0.73$^{+0.99}_{-0.28}$ & 7.6$^{+10.3}_{-2.9}$ & Ne\,\textsc{x} $1s$--$2p$ (1021.50) & $-$($290^{+250}_{-90}$) & 24.7 \\
1024.56$^{+0.18}_{-0.22}$ & 0.53$^{+0.17}_{-0.11}$ & 350$^{+120}_{-71}$ & $-$(1.48$^{+0.31}_{-0.17}$) & $-$(15.6$^{+3.2}_{-1.8}$) & Ne\,\textsc{x} $1s$--$2p$ (1021.50) & $-$($900^{+50}_{-60}$) & 129.3 \\
1055.91$^{+0.21}_{-0.20}$ & $< 0.63$ & $< 410$ & $-$(0.73$^{+0.20}_{-0.23}$) & $-$(7.0$^{+2.0}_{-2.2}$) & Fe\,\textsc{xxii} $2p$--$3d$ (1053.62) & $-$($650 \pm 60$) & 35.9 \\
1129.91$^{+0.96}_{-0.95}$ & 1.17$^{+0.80}_{-0.51}$ & 720$^{+490}_{-310}$ & $-$(0.59$^{+0.41}_{-0.24}$) & $-$(4.9$^{+3.4}_{-2.0}$) & Fe\,\textsc{xxiii} $2s$--$3p$ (1127.20) & $-$($720 \pm 250$) & 25.6 \\
1354.80$^{+0.78}_{-0.66}$ & 0.87$^{+0.84}_{-0.49}$ & 450$^{+430}_{-250}$ & $-$(0.74$^{+0.20}_{-0.59}$) & $-$(4.0$^{+1.1}_{-3.2}$) & Mg\,\textsc{xi} $1s$--$2p$ (1352.25) & $-$($570^{+170}_{-150}$) & 20.6 \\
1468.07$^{+1.19}_{-0.71}$ & $< 2.31$ & $< 750$ & 0.65$^{+0.33}_{-0.26}$ & 3.0$^{+1.5}_{-1.2}$ & Mg\,\textsc{xii} $1s$--$2p$ (1471.69) & 740$^{+150}_{-240}$ & 18.4 \\
1475.94$^{+0.23}_{-0.09}$ & $< 0.92$ & $< 430$ & $-$(0.81$^{+0.16}_{-0.24}$) & $-$(3.7$^{+0.8}_{-1.1}$) & Mg\,\textsc{xii} $1s$--$2p$ (1471.69) & $-$($860^{+50}_{-20}$) & 41.0 \\
1841.75$^{+1.56}_{-1.61}$ & 2.66$^{+2.11}_{-1.20}$ & 1000$^{+790}_{-450}$ & 1.93$^{+0.91}_{-0.62}$ & 5.5$^{+2.6}_{-1.8}$ & [Si\,\textsc{xiii}] $1s$--$2s$ (1839.42) & $-$($380^{+250}_{-260}$) & 40.7 \\
1869.17$^{+0.57}_{-0.44}$ & $< 1.06$ & $< 400$ & $-$(1.39$^{+0.33}_{-0.31}$) & $-$($3.9 \pm 0.9$) & Si\,\textsc{xiii} $1s$--$2p$ (1864.98) &$-$($670^{+90}_{-70}$) & 40.8 \\
\hline
\end{tabular}
\end{table*}

\begin{table*}
\centering
\caption{Table showing the spectral parameters in the rest frame of the best-fitting model simultaneously fit to the 2005 (obs.\,1) and 2008 (obs.\,2 and obs.\,3) broad-band 0.6--50.0\,keV {\sl Suzaku} data described in Section 6.5. $^{a}$Primary power-law continuum: $\Gamma$, photon index; Normalisation. $^{b}$Absorbed power-law. $^{c}$Partial-coverer: $N_{\rm H}$, column density; $\xi$, ionisation parameter; $f_{\rm cov}$, covering fraction expressed as a percentage. $^{d}$Neutral reflector: $A_{\rm Fe}$, iron abundance with respect to Solar. $^{e}$Zone of absorption. The redshift of the zone was fixed at the same value as that of the host galaxy (i.e. $z = 0.002336$). $^{f}$Ionised reflector. $^{g}$Highly ionised absorption: $v_{\rm out}$, velocity shift. $^{h}$Fe\,K$\alpha$ emission: $E_{\rm line}$, centroid energy; $\sigma$, intrinsic width; $F_{\rm line}$, line flux; EW, equivalent width. The symbol $p$ signifies that the parameter has pegged at the maximum / minimum value allowed by the model and the symbol $l$ signifies that the parameter is linked to all parameters with the corresponding number. Note that the Fe\,K$\alpha$ emission line parameters were obtained by replacing the neutral \textsc{reflionx} component with a \textsc{pexrav} component and a Gaussian.}
\begin{tabular}{l c c c c}
\hline\hline
& & \multicolumn{3}{c}{Value} \\
Component & Parameter & Obs.\,1 (2005) & Obs.\,2 (2008) & Obs.\,3 (2008) \\
\hline
\multirow{2}{*}{Power-law$^{a}$} & $\Gamma$ & $2.49^{+0.02^{l_{\rm 1}}}_{-0.01}$ & $l_{\rm 1}$ & $l_{\rm 1}$ \\
& Norm.\,(ph\,cm$^{-2}$\,s$^{-1}$) & $(1.72^{+0.03}_{-0.04}) \times 10^{-3}$ & $(1.23\pm0.02) \times 10^{-2}$ & $(8.11^{+0.07}_{-0.14}) \times 10^{-3}$ \\
\hline
\multirow{2}{*}{Power-law$^{b}_{\rm abs.}$} & $\Gamma$ & $l_{\rm 1}$ & $l_{\rm 1}$ & $l_{\rm 1}$ \\
& Norm.\,(ph\,cm$^{-2}$\,s$^{-1}$) & $(4.25^{+0.11}_{-0.08}) \times 10^{-3}$ & $(6.39^{+0.32}_{-0.30}) \times 10^{-3}$ & $(5.13^{+0.29}_{-0.28}) \times 10^{-3}$ \\
\\
\multirow{3}{*}{Partial-Coverer$^{c}$} & $N_{\rm H}$\,(cm$^{-2}$) & $(1.18^{+0.08}_{-0.09}) \times 10^{23^{l_{\rm 2}}}$ & $l_{\rm 2}$ & $l_{\rm 2}$ \\
& log\,$\xi$ & $0.51^{+0.11^{l_{\rm 3}}}_{-0.16}$ & $l_{\rm 3}$ & $l_{\rm 3}$ \\
& $f_{\rm cov}$ & $\sim$70\% & $\sim$30\% & $\sim$40\% \\
\hline
\multirow{4}{*}{\textsc{reflionx}$^{d}_{\rm abs.}$} & $\Gamma$ & $l_{\rm 1}$ & $l_{\rm 1}$ & $l_{\rm 1}$ \\
& $\xi$\,(erg\,cm\,s$^{-1}$) & $10^{^{l_{\rm 4}}}_{p}$ & $l_{\rm 4}$ & $l_{\rm 4}$ \\
& $A_{\rm Fe}$ & $0.27^{+0.03^{l_{\rm 5}}}_{-0.02}$ & $l_{\rm 5}$ & $l_{\rm 5}$ \\
& Norm.\,(ph\,cm$^{-2}$\,s$^{-1}$) & $(6.10^{+0.19}_{-0.28}) \times 10^{-5}$ & $(8.45^{+0.32}_{-0.62}) \times 10^{-5}$ & $(7.49^{+0.35}_{-0.44}) \times 10^{-5}$ \\
\\
\multirow{2}{*}{Absorber$^{e}$} & $N_{\rm H}$\,(cm$^{-2}$) & $(8.36^{+2.27}_{-1.75}) \times 10^{22^{l_{\rm 6}}}$ & $l_{\rm 6}$ & $l_{\rm 6}$ \\
& log\,$\xi$ & $0.60^{+0.24^{l_{\rm 7}}}_{-0.30}$ & $l_{\rm 7}$ & $l_{\rm 7}$ \\
\hline
\multirow{4}{*}{\textsc{reflionx}$^{f}_{\rm abs.}$} & $\Gamma$ & $l_{\rm 1}$ & $l_{\rm 1}$ & $l_{\rm 1}$ \\
& log\,$\xi$ & $3.5\pm0.1^{^{l_{\rm 8}}}$ & $l_{\rm 8}$ & $l_{\rm 8}$ \\
& $A_{\rm Fe}$ & $l_{\rm 5}$ & $l_{\rm 5}$ & $l_{\rm 5}$ \\
& Norm.\,(ph\,cm$^{-2}$\,s$^{-1}$) & $(3.67^{+1.00}_{-1.10}) \times 10^{-7}$ & $(9.78^{+1.48}_{-1.42}) \times 10^{-7}$ & $(8.68^{+2.08}_{-1.81}) \times 10^{-7}$ \\
\\
\multirow{2}{*}{Absorber$^{e}$} & $N_{\rm H}$\,(cm$^{-2}$) & $(4.84^{+0.16\,p}_{-0.17}) \times 10^{24^{l_{\rm 9}}}$ & $l_{\rm 9}$ & $l_{\rm 9}$ \\
& log\,$\xi$ & $2.53\pm0.02^{^{l_{\rm 10}}}$ & $l_{\rm 10}$ & $l_{\rm 10}$ \\
\hline
\multirow{3}{*}{High $\xi$ abs.$^{g}$} & $N_{\rm H}$\,(cm$^{-2}$) & $(5.49^{+0.71}_{-0.49\,p}) \times 10^{22^{l_{\rm 11}}}$ & $l_{\rm 11}$ & $l_{\rm 11}$ \\
& log\,$\xi$ & $3.9\pm0.1^{^{l_{\rm 12}}}$ & $l_{\rm 12}$ & $l_{\rm 12}$ \\
& $v_{\rm out}$\,(km\,s$^{-1}$) & $-(5\,400^{+600}_{-900})^{^{l_{\rm 13}}}$ & $l_{\rm 13}$ & $l_{\rm 13}$ \\
\hline
\multirow{4}{*}{Fe\,K$\alpha^{h}$} & $E_{\rm line}$\,(keV) & $6.40^{+0.01}_{-0.02}$ & $6.40^{+0.02}_{-0.01}$ & $6.41\pm0.02$ \\
& $\sigma$\,(eV) & $71^{+19}_{-23}$ & $68\pm33$ & $<55.3$ \\
& $F_{\rm line}$\,(ph\,cm$^{-2}$\,s$^{-1}$) & $2.06^{+0.22}_{-0.26}$ & $2.46^{+0.41}_{-0.36}$ & $1.70^{+0.43}_{-0.28}$ \\
& EW\,(eV) & $201^{+22}_{-25}$ & $102^{+17}_{-15}$ & $93^{+24}_{-15}$ \\
\hline
\end{tabular}
\end{table*}

\end{document}